\documentclass{JHEP3}

\title{  Poincare  field theory for  massless  particles }

\author{B. Sazdovi\'c \\
        Institute of Physics, P.O.Box 57, 11001 Belgrade, Serbia \\
        E-mail: \email{sazdovic@phy.bg.ac.rs}}

\abstract{

Our main proposition is that the field equations for all spins can be derived from the Casimir eigenvalue equations of the Poincaré group.   We have already confirmed this claim for massive scalar, spinor, and vector fields in Ref.\cite{S}. In this article, we extend this analysis to massless fields—specifically, the vector field, as well as second  and fourth  rank tensor fields  and finally for arbitrary rank tensor  fields.  In particular, we will derive the Maxwell equations,  the Einstein equations in the weak-field approximation as well as  Fronsdal equation.

As is well known,    defined particles as irreducible representations of the Poincaré group   \cite{Wi, BW}. However, as Weinberg noted in  \cite{W}, irreducible representations for massless vector fields with helicities $\pm 1$  do not exist. In this article, we confirm this statement for a broad class of massless fields. Unlike their massive counterparts, these fields are not Lorentz invariant because their Lorentz transformations include an additional gauge transformation term.  
Consequently, such fields can only appear in the theory in forms that are independent of the corresponding gauge parameters. These gauge invariant forms constitute our equations of motion. Thus, the massless case differs fundamentally from the massive one.

Finally, we demonstrate that our approach not only reproduces key results from well-known studies on highest     helicity fields but also extends to other helicities.

}

\usepackage{amsmath}

\begin{document}

\section{Introduction  }
\setcounter{equation}{0}

In our previous work  \cite{S}, we derived the equations of motion for massive particles of arbitrary spin by formulating them as eigenvalue equations for the Casimir operators of the Poincaré group.  
In this paper, we generalize this framework to massless particles, where new subtleties arise due to gauge invariance.

While only one Casimir operator (helicity) exists for massless particles, its covariant definition involves the 
Pauli-Lubanski vector  $W_a$,   
which possesses three independent components (due to the constraint 
 $P^a W_a = 0$). Thus, we obtain three equations, and their compatibility conditions lead to significant distinctions from the massive case.

As in the massive scenario, the fields are functions transforming under a representation of the Poincaré group. For a general representation, we denote these fields as   $\Psi^A (x)$, where  $A$ encompasses the relevant set of indices.

We will work in the standard momentum frame, which significantly simplifies our analysis.  
First, we solve the eigenvalue problem for the spin operator component  
$ (S_{1 2})^A{}_B = \sigma (W_0 )^A{}_B = ( W_3 )^A{}_B $. 
The resulting eigenvalues can be interpreted  as helicities 
 $\lambda_i$ and the corresponding eigenvectors    $\Psi^A_i $    
serve as candidates for irreducible representations.    
However, in certain  important  cases, the components of the Pauli-Lubanski vector fail to commute.
 Consequently, no common eigenvectors exist for all components.   
Moreover, in the standard momentum frame,   the operators  
 $W_1$  and  $W_2$  do not annihilate eigenvectors    $\Psi^A_i $, implying that     $\Psi^A_i $  
cannot represent genuine irreducible representations of the Poincaré group.

This feature has unexpected consequences, as discovered by Weinberg \cite{W}, where the Lorentz transformations of certain physically relevant fields (such as vector and symmetric second-rank tensor fields) acquire an additional term in the form of a gauge transformation. Consequently, for these fields, we can construct actions as gauge-invariant expressions involving the eigenvectors    $\Psi^A_i $. 

Thus, gauge-dependent fields are not Lorentz invariant and therefore do not constitute proper representations of the Poincaré group. However, the corresponding field strengths and Lagrangians—being gauge-invariant expressions of the gauge fields—do form representations of the Poincaré group. For this reason, we introduce the term incomplete representation of the Poincaré group (up to gauge transformations) for gauge fields, and incomplete irreducible representation of the Poincaré group for their corresponding irreducible components.

As a consequence, Wigner's definition of particles  \cite{Wi, BW} 
 as irreducible representations of the Poincaré group, while valid for massive cases, requires refinement in the massless case. Instead, we can define physical massless fields as equivalence classes, where two fields are considered equivalent if they are related by a gauge transformation. The action is then an expression that depends only on the equivalence class.

The main objective of this work is to demonstrate  that principle field equations  give rise to  all known field equations.  To this end, we first derive the standard equations for helicities  $1$  and $2$.
There is also  considerable  interest in massless fields with higher spin.
The theory of  particles with highest  spin was  originally  developed by Fierz and Pauli   \cite{FP}.
In Ref. \cite{Wi} and  \cite{BW} , it was  shown  that the one-particle states correspond to irreducible, unitary representations of the Poincaré group.

A minimal set of auxiliary traceless, symmetric tensor fields was later proposed in Ref.\cite{SH}. 
Since having an action is essential for field quantization, the equations of motion and corresponding action were derived in Ref.\cite{Fr}.
Another notable approach, based on a hierarchy of generalized Christoffel symbols, was introduced by de Wit and Freedman    \cite{dWF}.
All the above studies rely on the Fierz-Pauli constraints   \cite{FP}  which serve as their foundation. 
These constraints require that the field be a fully symmetric tensor of rank  $n$, 
while also being traceless, divergence-free, and satisfying the Klein-Gordon equation.

In this work, we present a more fundamental approach to analyzing massless tensor fields. 
 We develop a systematic procedure to determine all possible helicity states and the corresponding tensor structures for any given helicity.

As a special case, for a rank $n$ tensor field, the highest helicity state corresponds to a fully symmetric, traceless, and divergence-free tensor that satisfies the Klein-Gordon equation. Thus, our highest helicity states naturally satisfy the Fierz-Pauli constraints  \cite{FP}.  
Beyond the highest helicity, we also investigate   non-maximal helicity states and provide their physical interpretation (see Sections 7 and 8  and 12).

\section{Principle field equations}
\setcounter{equation}{0}

In this section, we introduce the fundamental field equations of the theory. These equations are constructed solely from the symmetry group properties and are formulated using the Casimir invariants.

In the present work, we begin with the massless Poincaré group and derive fundamental field equations that hold for arbitrary helicity states.

\subsection{Representation  for  Pauli-Lubanski vector }

Casimir invariants are  expressions that commute with all group elements.
In   massless case,  where square of momentum vanishes  $P^2 = 0$,   there are   two Casimir operators for Poincare group,   the  helicity  $\lambda$ and   sign of $P^0$.
The covariant  definition of helicity is
\begin{eqnarray}\label{eq:COP10}
 W_a = \lambda P_a       \,  ,  \qquad  W_a = \frac{1}{2} \varepsilon_{a b c d} S^{b c} P^d     \, ,
\end{eqnarray}
where $W_a$  is  Pauli-Lubanski vector and   $S_{a b}$ is spin parts of Lorentz  generators.
Since   $P^a W_a = 0$ we can conclude  from  (\ref{eq:COP10})    that $P^2 = 0$, as it should be for massless case.

Two comments should be made. First, it is possible to  use  non covariant  definition of helicity $\lambda =  S_i n_i $   as  projection of  space part of
spin generator  $S_i =  \frac{1}{2} \, \varepsilon_{i j k}  S_{j k}$  to  momentum axis   $ n_i  =  \frac{ p_i }{ |p_i |^2  }$.
It produces  the same spectrum  for  $\lambda$ as relation  (\ref{eq:COP10}).   But in that case we  are losing the crucial
possibility to  insist on Lorentz invariance which will product gauge transformations.
Second, in definition of  Pauli-Lubanski vector  orbital part of four dimensional rotation generatr  $L_{a b} =    x_a P_b - x_b P_a $  does not appear  since $\varepsilon_{a b c d} L^{b c} P^d = 0 $.

For all types of fields representation of momentum is   $(P_a)^A{}_B  \to   \delta^A_B \, i \partial_a$. In  article  \cite{S}  we proved relation
\begin{eqnarray}\label{eq:BEder}
 (S_{a b})^{A B}{}_{C D}    =  (S_{a b})^{A}{}_{C} \delta_D^B  +  \delta^A_C (S_{a b})^{B}{}_{D}  \, ,
\end{eqnarray}
which allows us to find  representations for arbitrary spin operator   using fundamental  representations.
This initial  representation  for   Dirac spinors  and  vector fields are
\begin{eqnarray}\label{eq:gLr21}
(S_{a b})^\alpha{}_\beta   = \frac{i}{4} [\gamma_a, \gamma_b ]^\alpha{}_\beta   \, ,     \qquad     (S_{a b})^c{}_d  =  i \Big(    \delta^c_a   \eta_{b d} - \delta^c_b   \eta_{a d}  \Big)         \, .
\end{eqnarray}
For example, the expression for vectors in terms of spinors  $ V^a (x) =  {\bar \psi} (x)  \gamma^a   \psi (x) $   and first relation   (\ref{eq:gLr21}) produces the second one, which  can also been  obtained by direct calculation.

Using representations for  momenta and for  spin generators  as well as  the fact that they commute, we can find representation for   Pauli-Lubanski vector.

\subsection{Standard momentum $k^a$,  space inversion and little group }

It is useful to work  in frame of  standard momentum $k^a$. In massless  case, when  Lorentz invariant function of momentum vanish,
$p^2 = 0$,  we can chose  standard momentum  in the form $k^a = (1, 0, 0, - 1  )$.  Then we can express any momentum  $p^a$  as Lorentz transformation of $k^a$
\begin{eqnarray}\label{eq:paLka}
p^a =  L^a{}_b  (p)  k^b \, .
\end{eqnarray}

Up to now we considered proper and orthochronous   Lorentz group $\Lambda  \in  SO (1, 3)^\uparrow$ which contains the identity matrix and preserves the direction of time and parity
( $\det \Lambda = 1 $  and $\Lambda^0{}_0 \ge 1$ ). But we will also  need   space  inversion   ( $\det \Lambda = - 1 $  and $\Lambda^0{}_0 \ge 1$ ) which for vector fields is
implemented by the operator
\begin{eqnarray}\label{eq:}
{\cal P}^a{}_b  =  \delta^a_0 \delta_b^0 -   \delta^a_i \delta_b^i   \, .  \qquad  (i= 1, 2, 3)
\end{eqnarray}
In order to follow behavior under space  inversion  we will introduce new parameter
$ \sigma $     with  inversion rules  ${\cal P} \sigma =  -  \sigma  $,  and obviously  $\sigma^2 = 1$.  Then
rotations are invariant under parity  $S_{i j }  \to S_{i j }  $  and  boosts change the sign   $S_{0 i }  \to \sigma S_{0 i }  $  while   $k^a \to  (1, 0, 0, - \sigma  )$.

Since  $P^a W_a  = 0$,   Pauli-Lubanski vector has three independent  components and for standard momentum  they are
\begin{eqnarray}\label{eq:PLLg10}
W_0 = \sigma  S_{1 2 } \, ,  \qquad  W_3 =   S_{1 2 } \, ,  \qquad   \nonumber \\
W_1  \equiv    \Pi_2   =   S_{0 2} - S_{3 2}   \, , \qquad
- W_2    \equiv   \Pi_1   =  S_{0 1} -    S_{3 1}   \, .
\end{eqnarray}
All three generators annihilate standard momentum.  So  group element  $W^a{}_b  $,  leaves  $k^a$ invariant  $ W^a{}_b k^b = k^a $,
which is definition of  little group in massless Poincare case.

Commutation relations of little group generators  have the  form
\begin{eqnarray}\label{eq:}
[\Pi_1, \Pi_2] =  0     \,  ,   \qquad    [S_{1 2}, \Pi_1] = i \Pi_2    \,  ,   \qquad    [S_{1 2}, \Pi_2] = - i \Pi_1        \,  .
\end{eqnarray}
Therefore,  as it is well known,  in massless case the little group of Poincare group   is   group $E (2)$, which consists of translations and rotations in two-dimensional Euclidean space.

Note that unlike the massive case here little group contains Abelian subgroup with generators $\Pi_1$  and  $\Pi_2$.
Representation of little group  must be invariant under this Abelian subgroup.  We will see that violation of  this feature has important  repercussion   that gauge invariance is  consequence of Lorentz transformation   \cite{W}.

\subsection{Principle field  equations  }

Using the expression    (\ref{eq:COP10})  along with the momentum    and   Pauli-Lubanski vector   we  postulate  {\it  principle field  equations for massless  fields   $\Psi^A (x)$  with    helicity  $\lambda$ }
\begin{eqnarray}\label{eq:EMmsc2}
  (W_a)^A{}_B   \Psi^B (x)  =  \lambda (P_a)^A{}_B   \Psi^B (x)          \, .
\end{eqnarray}
We demonstrate   that  all massless free field equations with  arbitrary helicity   in classical  field theory   \cite{W, BS, Pes, LL, WG}   can be  derived   from this   fundamental  equation. 
In this work, we explicitly verify this claim for the most physically significant massless fields.

We will  investigate principle field   equations in momentum representation  in the frame of standard momentum.
Then  differential equations  (\ref{eq:EMmsc2})   become algebraic equations.
After solving this algebraic equations we can go back to  $p^a$ dependent solutions using  (\ref{eq:paLka})  and then to coordinate representation.

Using  (\ref{eq:EMmsc2})  and (\ref{eq:PLLg10})    we can obtain  principle  field equations for standard momentum
\begin{eqnarray}\label{eq:SGcom}
   (S_{1 2})^A{}_B  \Psi^B (k)   =  \lambda  \sigma  \Psi^A (k)         \, ,   \nonumber \\
 (\Pi_1 )^A{}_B \Psi^B (k) = 0    \, , \qquad     (\Pi_2 )^A{}_B \Psi^B (k) = 0    \,  .
\end{eqnarray}

\subsection{Consequences of parity conservation  }

The  operators $(S_{1 2})^A{}_B$,  $ (\Pi_1 )^A{}_B$ and $ (\Pi_2 )^A{}_B$ are  representation of  little group  generators  and
$\lambda  \sigma$ is a eigenvalue of operator $ (S_{1 2})^A{}_B$.  Under space inversion  operator $ (S_{1 2})^A{}_B$  is invariant  which  means that   helicity  change the sign
\begin{eqnarray}\label{eq:}
{\cal P} \lambda  =  -  \lambda   \,  .
\end{eqnarray}
Therefore,  if the theory is invariant under space inversion then for any non zero helicity it must exist opposite one.

Helicity is invariant under  proper orthochronous Poincare  group.  Particles of opposite helicities   $+ \lambda$   and $- \lambda$,
which are related by transformation of space inversion,   corresponds to different species of particles.
A  theory that conserves parity must treat both helicities   states equally. So,  both massless states   with  opposite helicities we can consider as two polarizations of the same particle.
This particle  is  irreducible representation  of wider group which combined  Poincare group and parity.

The  most important examples are electromagnetic and gravitation  interactions which are invariant  under space inversion.  In particular,  both massless   particles
with helicities $\lambda = 1$ and $\lambda = - 1$  are called photons  and  both massless particles  with helicities $\lambda =  2$  and  $\lambda = - 2$ are called  gravitons.
General state of massless particle which includes both helicities $ \lambda $  and  $-  \lambda $  is
\begin{eqnarray}\label{eq:GSbh}
\Psi^A  (x)  =   \alpha   \Psi^A_{+ \lambda} (x)  +  \beta  \Psi^A_{- \lambda}  (x)  \, ,  \qquad    | \alpha |^2  +   | \beta |^2  = 1  \,  ,
\end{eqnarray}
and   in momentum representation
\begin{eqnarray}\label{eq:}
 \Psi^A  ( p)  =     \alpha   \Psi^A_{+ \lambda} (p)  +  \beta  \Psi^A_{- \lambda}   ( p)   \, .
\end{eqnarray}

\subsection{Real fields  in coordinate and momentum  representation  }

In most cases we will work  with  real fields,  such that in coordinate and momentum  representation
\begin{eqnarray}\label{eq:Ref}
(\Psi^A)^\ast (x) =  \Psi^A (x) \, ,  \qquad  (\Psi^A)^\ast (p) =  \Psi^A ( - p)  \, .
\end{eqnarray}
So for real fields
\begin{eqnarray}\label{eq:Ref1}
 \Psi^A (x) =   \int d^4 p  \Big( \Psi^A  (p) +  \Psi^A ( - p)  \Big)  e^{- i p x}   \, .
\end{eqnarray}
Similarly,  for  imaginary field we have
\begin{eqnarray}\label{eq:Imf}
(\Psi^A)^\ast (x) = - \Psi^A (x) \, ,  \qquad  (\Psi^A)^\ast (p) = - \Psi^A ( - p)  \, .
\end{eqnarray}

We will need real part  of the expression which in momentum   representation  has a form  $\Psi_n^B  (p)  =   p^{a_1} p^{a_2} \cdots p^{a_n}  \Psi^B (p) $
\begin{eqnarray}\label{eq:}
\Psi_n^B (x) =  \int  d^4 p  \Big[  \Psi_n^B  (p) +  \Psi_n^B  (- p)   \Big]  e^{- i p x}
=  (i)^n   \partial^{a_1} \partial^{a_2} \cdots \partial^{a_n} \int  d^4 p  \Big[  \Psi^B  (p) +  (- 1)^n    \Psi^B (- p)  \Big]  e^{- i p x}    \, .  \qquad
\end{eqnarray}
For $n = 1$ we have
\begin{eqnarray}\label{eq:Om1cr}
\Psi_1^B (x) =   \int  d^4 p  \Big[  p^a \Psi^B (p) - p^a   \Psi^B (- p)  \Big]  e^{- i p x}
 =   \partial^a  \Omega_1^B  (x) \, ,  \qquad  \qquad
\end{eqnarray}
and for $n = 2$
\begin{eqnarray}\label{eq:Om2cr}
\Psi_2^B (x) =   \int  d^4 p  \Big[  p^a p^b  \Psi^B  (p) +  p^a p^b  \Psi^B (- p)  \Big]  e^{- i p x}
=   \partial^a \partial^b \Omega_2^B  (x)  \, ,
\end{eqnarray}
where
\begin{eqnarray}\label{eq:om1x}
 \Omega_1^B  (x)  =  i    \int  d^4 p  \Big[ \Psi^B  (p) -  \Psi^B  (- p)  \Big]  e^{- i p x}    \, ,  \qquad
\Omega_2^B  (x)  =   -  \int  d^4 p  \Big[ \Psi^B  (p) +   \Psi^B  (- p)  \Big]  e^{- i p x}  \, .  \qquad \qquad
\end{eqnarray}
According to  (\ref{eq:Imf})  and  (\ref{eq:Ref})  both $ \Omega_1^B  (x)$  and  $ \Omega_2^B  (x)$  are real functions.

\subsection{Solution of $(S_{1 2})^A{}_B$    equation }

We are going to  solve eigenproblem  of   operator   $(S_{1 2})^A{}_B$ which  will produces spectrum of   helicities $\lambda_i$  $(i = 1, 2, \cdots , n)$.  Then we will  find
projection operators   $( P_i )^A{}_B$   and   eigenfunctions  $\Psi_i^A$ corresponding to these eigenvalues.

In order that first  equation   (\ref{eq:SGcom})    has nontrivial solutions    its characteristic   polynomial must vanish
\begin{eqnarray}\label{eq:DetS1}
 P_{S_{1 2}} (\lambda \sigma  )  =  \det \Big(   (S_{1 2})^A{}_B -  \lambda  \sigma \delta^A_B   \Big)  =  0   \, .
\end{eqnarray}
The zeros of  characteristic   polynomial are  eigenvalues $\lambda_i$.

Next,   we can construct   projection operators  $(P_i)^A{}_B$ on subspaces  $\Xi_i$  (we will call it eigenspace)   with dimension $d_i = dim \Xi_i$  corresponding to  helicities $\lambda_i$
\begin{eqnarray}\label{eq:PiAB}
( P_i )^A{}_B
=   \frac{  \Big[  \prod_{j \neq i}^n    \Big(  S_{1 2}   - \lambda_j \sigma  \delta  \Big) \Big]^A{}_B }{  \prod_{j \neq i}^n    \Big(  \lambda_i \sigma   - \lambda_j \sigma  \Big) }          \, ,
\end{eqnarray}
and obtain corresponding  eigenfunctions
 \begin{eqnarray}\label{eq:IrrPa}
\Psi_i^A  (k) = ( P_i )^A{}_B \Psi^B   (k)    \, .
\end{eqnarray}

From    (\ref{eq:PiAB})   and    (\ref{eq:IrrPa})   it follows that  space inversion connect projectors  and eigenfunctions  with opposite helicities
\begin{eqnarray}\label{eq:PiAB1}
{\cal P} \,  ( P_{\lambda_i} )^A{}_B  =  ( P_{- \lambda_i} )^A{}_B    \, ,   \qquad
{\cal P} \, \Psi_{\lambda_i}^A  (k)       =    \Psi_{ - \lambda_i}^A  (k)        \, .
\end{eqnarray}

Generally, (see section   \ref{sec:Tar} for details)    with the help of projection operators,  for tensor fields of rank $n$  we can rewrite equation   (\ref{eq:DetS1})    in the form
\begin{eqnarray}\label{eq:Smlade}
\det ( S_{1 2} - \lambda \sigma \delta)
=     \prod_{i = - n}^n   (i - \lambda \sigma)^{\tilde{d}_i}     \det  P_i    = 0         \,   ,    \qquad
\end{eqnarray}
and using $\sigma^2 = 1$  and $\tilde{d}_i = \tilde{d}_{- i}$
\begin{eqnarray}\label{eq:Smlade1}
\det  ( S_{1 2} - \lambda \sigma  \delta)
=   ( - \lambda \sigma)^{\tilde{d}_0}   \prod_{i =1}^n   ( i^2 - \lambda^2 )^{\tilde{d}_i}   \, \,
 \prod_{i = - n}^n   \det  P_i      = 0         \,   .    \qquad
\end{eqnarray}
Exponents in multipliers  $\tilde{d}_i$ are algebraic  multiplicity while  $d_i$ as   dimensions of eigenspaces   $\Xi_i$   are   geometric multiplicity. Generally, $d_i \leq \tilde{d}_i $, but in our case
$d_i = \tilde{d}_i $  so that  from now on we will use a unique notation $d_i$.

Note that  for tensor fields of rank  $n$   relation
$\sum_{i = - n}^n  d_i  =   d_0 + 2 \sum_{i = 1}^n  d_i  =  4^n $  is satisfied and   spectrum of  helicities is
\begin{eqnarray}\label{eq:}
\lambda_0 = 0 \, , \qquad     \lambda_{\pm  i} = \pm i     \, ,   \qquad  i = (1, 2, \cdots, n )    \,  .
\end{eqnarray}

Let us stress that gauge invariant  fields $\Psi_i^A  (k)$  (see the next subsection for explanation)  are representations of Poincare group,
but in general case  they are  not irreducible representations. To get  irreducible representations we must separate fields
$\Psi_i^A  (k)$  using symmetry properties.
Note that  dimension of subspace  $d_i = dim \Xi_i$  corresponding to  helicity $\lambda_i$  is  in fact number of degrees of freedom of  fields   $\Psi_i^A  (k)$.
Similarly,  dimension of subspace of irreducible representation is  number of degrees of freedom of corresponding  fields.
We are going to check it in all  examples.

\subsection{Lorentz transformations induce  gauge transformations  }

We solved first equation (\ref{eq:SGcom}) and it produces complete set of  eigenvalues and  eigenfunctions. But we have additional two conditions.
It is of great importance to investigate  the  compatibility of all  equations (\ref{eq:SGcom}).

\subsubsection{Gauge dependent fields }

In equations for standard momentum     (\ref{eq:SGcom})   operators $\Pi_1$  and $\Pi_2$ should  annihilate field $\Psi^A (k)$.  But explicit calculation shows that this does not happen
in some physically relevant cases.  Since equations (\ref{eq:SGcom})  are  condition for Lorentz invariance,   violation of these conditions breaks  Lorentz invariance.
To measure this violation  we introduce expression
\begin{eqnarray}\label{eq:Txyvg}
\delta  \Psi_i^A (\varepsilon_1, \varepsilon_2 ) (k)    =  i \Big(  \varepsilon_1 \Pi_1  +  \varepsilon_2 \Pi_2  \Big)^A{}_B    \Psi_i^B (k)    \, ,   \qquad
\end{eqnarray}
where $ \varepsilon_1$  and $ \varepsilon_2$ are some parameters.

Field equations  must be  Lorentz invariant and consequently they  should not depend on $\delta  \Psi_i^A (\varepsilon_1, \varepsilon_2 ) (k)$. It means that
\begin{eqnarray}\label{eq:}
\delta  \Psi_i^A (\varepsilon_1, \varepsilon_2 ) (k)   =  {\cal G}_i  \Big( k, \Omega_\pm  \Big)  \, ,   \qquad   \Omega_\pm =  \Omega_\pm (\varepsilon_\pm)
\end{eqnarray}
are  gauge transformations of fields $ \Psi_i^A (k) $  with gauge parameters $\Omega_\pm (\varepsilon_\pm)$.  So, as Weinberg  noted  Lorentz transformations induce  gauge  transformations  \cite{W}.

We can  perform  transition from standard momentum $k^a $ to arbitrary one $p^a$    and using relation  $p^a = i \partial^a$  we can  obtain   gauge transformation in coordinate space
\begin{eqnarray}\label{eq:}
\delta    \Psi_i^A  (\Omega_\pm, x )  =   {\cal G}_i  \Big( i \partial, \Omega_\pm  (x) \Big)         \, .
\end{eqnarray}
Finally, we can chose the action
\begin{eqnarray}\label{eq:}
S (\Psi_i) =  \int d^4 x {\cal L} \Big( \Psi_i^A (x) \Big) \, ,
\end{eqnarray}
as a   gauge invariant expressions of  $ \Psi_i^A (x) $,  which means that following condition must be satisfied
\begin{eqnarray}\label{eq:}
S \Big( \Psi_i +  \delta  \Psi_i  \Big)   =    S \Big( \Psi_i \Big)  \, .
\end{eqnarray}

Fields which  violate Lorentz invariance  are  gauge dependent fields  and   we  will   simply call them  gauge fields.  They  are  not representation of Poincare group.  If  we want to weaken the statement
they  are   representation of Poincare group    up to gauge transformations,   since  corresponding  field strength and  Lagrangian
are  representations of Poincare group.  Therefore, for these fields  we will introduce  name   {\it incomplete} representation of Poincare group.

\subsubsection{Gauge invariant  fields }

Gauge invariant  fields are   representation of Poincare group. So, they  are physical and we can use them without restrictions. We can obtain their  form  as solutions of equations  (\ref{eq:SGcom}).
That leads us to  field equations  of  gauge invariant  fields as  some constraints on the obtained solution.

Next   we will consider the  most important  physical cases of massless fields: scalar field,  vector field with  helicities $\lambda= \pm 1$ (photon) and   symmetric second rank  tensor   with  helicities  $\lambda= \pm 2$   (metric). We will also consider second rank  tensor  with  helicities $\lambda= \pm 1$  (field strength) and  fourth  rank  tensor   with  helicities $\lambda= \pm 2$  (Riemann curvature).
We will also comment gauge fields with higher helicities.

\section{Scalar  field   }
\setcounter{equation}{0}

The simplest  example  is scalar field $\varphi (x)$. In that case spin operator   $S_{a b }$  vanishes.   In particular
$S_{1 2 } = 0$ which means that helicity is zero, $\lambda = 0$. From   $\Pi_1 = 0  =  \Pi_2$  we obtain    $\delta  \varphi (\varepsilon_1, \varepsilon_2 ) (x) = 0$.  Therefore,  massless scalar field $\varphi (x)$ is gauge invariant, and we can take it as physical field without restrictions.   It  is one dimensional irreducible representation of Poinace group   and consequently  it has  one degree of freedom.
Equation of motion is   massless  Klein-Gordon equation $\partial^2 \varphi (x) = 0$.

\section{Vector field  and Maxwell equations  }
\setcounter{equation}{0}

The case of  massless vector field is most important.  Besides representing electromagnetic interaction  we will see  that  solutions for massless tensor fields of arbitrary rang
comes down to the case  of  vector field.

\subsection{Eigenvalues and  projection operators    }

For vector field  we have  $A, B \to a, b$,    $\Psi^A \to V^a$   and
\begin{eqnarray}\label{eq:SOvf}
(S_{a b})^A{}_B    \to  \Big( S_{a b}  \Big)^c{}_d = i \Big( \delta^c_a \, \eta_{b d}  - \delta^c_b \, \eta_{a d}  \Big)   \,  ,         \qquad    \Rightarrow  \qquad
( S_{1 2 } )^a{}_b  =    i \Big(    \delta^a_1   \eta_{2 b} - \delta^a_2   \eta_{1 b}  \Big)     \, .
\end{eqnarray}

The  consistency condition requires   that characteristic polynomial vanishes. It means
\begin{eqnarray}\label{eq:}
 \det \Big( {\cal S}_{1 2} - \lambda \sigma \Big)^a{}_b    =  \lambda^2 (\lambda - 1  ) (\lambda + 1  ) = 0    \, ,
\end{eqnarray}
 with solutions for eigenvalues
\begin{eqnarray}\label{eq:SEIV}
\lambda_0  =  0 \, ,   \qquad   \lambda_{\pm  1}  =  \pm 1 \, .  \qquad
\end{eqnarray}

We note that eigenspace $\Xi_0$  corresponding to  the eigenvalue  $\lambda_0 = 0$  is two dimensional     $d_0 = dim \Xi_0 = 2$  implying that its associated eigenvector carries two degrees of freedom.
Similarly, the  eigenspaces   $\Xi_{\pm 1}$   corresponding to eigenvalues $\lambda_{\pm 1} = 1$  
respectively are both  one dimensional   $d_{\pm 1}  =  dim \Xi_{\pm 1} = 1$. 
Consequently, each of their associated eigenvectors carries a single degree of freedom.

We already had general expression for projection operators   (\ref{eq:PiAB})     which for vector fields, where $n = 3$,  takes the  form
\begin{eqnarray}\label{eq:PiAB1v}
( P_i )^a{}_b  =    \frac{    \prod_{j \neq i}^3    \Big(  (S_{1 2})^a{}_b   - \lambda_j \sigma \delta^a_b  \Big) }{  \prod_{j \neq i}^3    \Big(  \lambda_i \sigma  - \lambda_j  \sigma \Big) }         \, ,
\end{eqnarray}
or explicitly
\begin{eqnarray}\label{eq:}
( P_{ 0} )^a{}_b (k)
=  \delta^a_b    -    (S_{1 2}^2)^a{}_b   =  \delta^a_b    -   \delta^a_\alpha \delta_b^\alpha    \, , \qquad  (\alpha = 1, 2 )     \nonumber \\
( P_{ \pm 1} )^a{}_b   (k)  = \frac{1}{2}    \Big(   (S_{1 2}^2)^a{}_b    \pm  \sigma (S_{1 2})^a{}_b \Big)
= \frac{1}{2}    \Big[  \delta^a_\alpha \delta_b^\alpha     \pm   i \sigma \Big(    \delta^a_1   \eta_{2 b} - \delta^a_2   \eta_{1 b}  \Big)    \Big]   \, .
\end{eqnarray}

\subsection{Basic vectors  }

For further benefits we introduce  basic vectors $k^a$  and $q^a$ in two dimensional eigenspace  $\Xi_0 $.  Then complete set of basic vectors
$e^a_i = \{k^a, q^a,  \breve{p}_+^a,  \breve{p}_-^a  \}$  where $ i=\{ 0+, 0-, +1, -1  \} $  are
\begin{eqnarray}\label{eq:kqp}
k^a   = \delta^a_0 - \sigma \delta^a_3   \, ,  \quad
q^a    =  \delta^a_0 + \sigma \delta^a_3  \, ,  \quad
 \breve{p}_\pm^a    =  \delta^a_1  \pm i \sigma   \delta^a_2      \, .  \qquad
\end{eqnarray}
It is easy to check that
\begin{eqnarray}\label{eq:}
(S_{1 2})^a{}_b e^b_i = \lambda_i e^a_i  \, ,
\end{eqnarray}
where $\lambda_i$  take the values from  (\ref{eq:SEIV}).  Consequently,  basic vector $e^a_i$   carries helicity $\lambda_i$.
In particular,  basic vectors $ \breve{p}_\pm^a$   carry  helicities  $\pm 1$ and both basic vectors $k^a$  and $q^a$   carry  helicities  $0$.
If $n_\pm$  are numbers of vectors  $ \breve{p}_\pm^a$  in the representation of field  $\Psi^A$,   then its helicity is
\begin{eqnarray}\label{eq:Hln}
\lambda = n_+ - n_-  \, .
\end{eqnarray}
Note that tensors with highest helicity (for example tensor of rank$n$ with helicities $\lambda_{\pm n}  = \pm n$)   have a form $\breve{p}_\pm^{a_1}  \breve{p}_\pm^{a_2} \cdots \breve{p}_\pm^{a_n} \mathbf{T} $.
Consequently,  tensors with highest helicity  are  one dimensional and they are symmetric in all indices.  We will confirm   that in all examples.

Projection operators in terms of  basic vectors take the form
\begin{eqnarray}\label{eq:PRkqp5}
( P_{ 0 +} )^a{}_b (k)  =  \frac{1}{2}   q^a  k_b  \, , \qquad   ( P_{ 0 -} )^a{}_b (k)  =  \frac{1}{2}   k^a  q_b  \, , \qquad
( P_{ \pm 1} )^a{}_b   (k)  = - \frac{1}{2}   \breve{p}_\pm^a   (\breve{p}_\mp)_b   \, .
\end{eqnarray}
We can check that above expressions   are really projectors  using  product of basic vectors where only nontrivial products  are
\begin{eqnarray}\label{eq:SPbv}
k^a q_a = 2  \, ,   \qquad   \breve{p}_\pm^a   (\breve{p}_\mp)_a  =  - 2    \, .
\end{eqnarray}
Note that all basic vectors are light-like.

\subsection{Eigenfunctions of operator $(S_{1 2})^a{}_b$  }

The eigenfunctions of operator $(S_{1 2})^a{}_b$     are $V_i^a  (k ) =  (P_i)^a{}_b V^b $  where  $ i=\{ 0+, 0-, +1, -1  \} $,   or explicitly
\begin{eqnarray}\label{eq:Vfbv}
V_{0 +}^a  (k )  =     \frac{1}{2}   q^a  k_b V^b    \equiv q^a \mathbf{V}_k   \, , \qquad
 V_{0 -}^a  (k ) =     \frac{1}{2}   k^a  q_b V^b  \equiv k^a \mathbf{V}_q   \, ,     \qquad   \nonumber \\
  V_{\pm 1}^a  (k )    =    - \frac{1}{2}    \breve{p}_\pm^a   (\breve{p}_\mp)_b   V^b  \equiv   \breve{p}_\pm^a  \mathbf{V}_\mp             \,  .   \qquad
\end{eqnarray}
It is easy to check that relation  (\ref{eq:Hln})  is satisfied  since scalars like  $\mathbf{V}_\mp = (\breve{p}_\mp)_b   V^b$  carry helicity $0$.

The basic vectors determines the characteristics of eigenvectors. They have the same helicity and as we will see the same  gauge transformations and the same  parity.
Here all eigenfunctions  $V_i^a  (k )$   are  potential representations of Poincare group, but only  gauge invariant ones are really  representations of Poincare group.

Under space inversion  vectors  $k^a$  and $q^a$ are invariant  while
\begin{eqnarray}\label{eq:}
{\cal P}  \breve{p}_\pm^a    =  -   \breve{p}_\mp^a  \, .
\end{eqnarray}
It confirms  that  general statements (\ref{eq:PiAB1}) valid  for vector fields
\begin{eqnarray}\label{eq:}
{\cal P}   ( P_{ \pm 1} )^a{}_b   (k)  =    ( P_{ \mp 1} )^a{}_b   (k)     \,  ,     \qquad
{\cal P}    V_{ \pm 1}^a (k)  =    V_{ \mp 1}^a     (k)   \, .
\end{eqnarray}

The vector field $V^a$ has four components. There are four  corresponding eigenfunctions  $V^a_{ i}$    with four components. But they are not independent.
In fact,  if we multiply all  components  $V^a_{ i}$ with all  basic vectors we obtain only four nonzero expressions
\begin{eqnarray}\label{eq:Scgi}
k_a V_{0 +}^a    =     k_b V^b  = 2 \mathbf{V}_k  \, , \qquad
q_a V_{0 -}^a   =      q_b V^b  =   2 \mathbf{V}_q \, ,     \qquad
(\breve{p}_\mp)_a  V_{\pm 1}^a   =   (\breve{p}_\mp)_b   V^b  = - 2 \mathbf{V}_\mp           \,  .   \qquad
\end{eqnarray}

\subsection{Gauge transformations   of  massless vector  fields }

Gauge transformations of eigenfunctions of operator $(S_{1 2})^a{}_b$  for  massless vector fields
\begin{eqnarray}\label{eq:Txyvgv}
\delta  V_i^a   ( \varepsilon_1, \varepsilon_2 )  (k)
=   i \Big(  \varepsilon_1 \Pi_1  +  \varepsilon_2 \Pi_2  \Big)^a{}_b    V_i^b  (k)   \, , \qquad
\end{eqnarray}
are  particular case of   Eq.(\ref{eq:Txyvg}). Explicitly,  using expressions
\begin{eqnarray}\label{eq:PLLg1}
( \Pi_1 )^a{}_b
=  i \Big(   \delta^a_0   \eta_{1 b} -  \delta^a_1   \eta_{0 b}  +   \delta^a_1   \eta_{3 b} - \delta^a_3   \eta_{1 b}  \Big)   \, , \qquad
  (  \Pi_2  )^a{}_b
  =  i \Big(   \delta^a_0   \eta_{2 b} -    \delta^a_2   \eta_{0 b}  +   \delta^a_2   \eta_{3 b} - \delta^a_3   \eta_{2 b}  \Big)     \, ,   \qquad
\end{eqnarray}
we  get
\begin{eqnarray}\label{eq:PLLg1n}
\delta  V_i^a   ( \varepsilon_1, \varepsilon_2 )
=  -  \Big[  \varepsilon_1  \Big(    \delta^a_0   \eta_{1 b} -  \delta^a_1   \eta_{0 b}  +   \delta^a_1   \eta_{3 b} - \delta^a_3   \eta_{1 b}  \Big)
 +   \varepsilon_2   \Big(     \delta^a_0   \eta_{2 b} -   \delta^a_2   \eta_{0 b}  +   \delta^a_2   \eta_{3 b} - \delta^a_3   \eta_{2 b}  \Big)   \Big]    V_i^b   \nonumber \\
 =  \left| {\begin{array}{cccc}
 0  &    \varepsilon_1    &   \varepsilon_2     &   0     \\
  \varepsilon_1   &  0   &  0   &   \varepsilon_1   \\
  \varepsilon_2    &  0  &  0  &   \varepsilon_2  \\
0     &  -  \varepsilon_1  &  -  \varepsilon_2   &   0    \\
\end{array} } \right|  \left| {\begin{array}{c}
   V^0_i     \\
      V^1_i   \\
    V^2_i    \\
      V^3_i  \\
 \end{array} } \right|  \, .   \qquad  \qquad   \qquad
\end{eqnarray}
In particular, the   gauge transformations of basic vectors  have simple form
\begin{eqnarray}\label{eq:gtbv}
\delta  k^a   = 0   \, ,  \qquad
\delta  q^a  =       \varepsilon_+   \breve{p}_-^a   +   \varepsilon_-   \breve{p}_+^a       \, ,      \qquad
 \delta  \breve{p}_\pm^a  =   \varepsilon_\pm   k^a    \, .  \qquad
( \varepsilon_\pm =  \varepsilon_1  \pm i \sigma  \varepsilon_2    )   \qquad
\end{eqnarray}

The four independent scalars (\ref{eq:Scgi})  are gauge invariant. So, gauge transformations of  eigenfunctions  $V_i^a  (k ) =  (P_i)^a{}_b V^b (k)$  are  determined by gauge transformations of basic vectors.

Let us stress that  only  component  $ V_{0 -}^a  (k )$ is gauge invariant,  $ \delta  V_{0 -}^b  (k ) =  0 $,   and consequently  only this part is irreducible representation of Poincae group.
The other components   $ V_{0 +}^a  (k )$  and  $V_{\pm 1}^b  (k )$  are not gauge invariant and    they transform as
\begin{eqnarray}\label{eq:GtrV}
\delta   V_{0 +}^a  (k )  =    \omega_+ \breve{p}_-^a  +      \omega_-  \breve{p}_+^a        \, ,  \qquad
  \omega_\pm   (k )   = \frac{1}{2}  k_a  V^a  \varepsilon_\pm     \, ,   \nonumber \\
\delta    V_{\pm 1}^a  (k )   =    k^a  \Omega_\pm      \, ,     \qquad
\Omega_\pm   (k ) =    - \frac{1}{2}   (\breve{p}_\mp)_b    V^b   \varepsilon_\pm      \, .
\end{eqnarray}

From  (\ref{eq:gtbv}) we can conclude that under space inversion  ${\cal P} \varepsilon_\pm = - \varepsilon_\mp$. It produces  ${\cal P} \omega_\pm = - \omega_\mp$  and
${\cal P} \Omega_\pm =  \Omega_\mp$.  It is in agreement with relations ${\cal P}  V_{0 +}^a  (k ) =  V_{0 +}^a  (k )$   and  ${\cal P}  V_\pm^a  (k ) =  V_\mp^a  (k )$.

\subsection{Massless vector  field  with helicity  $\lambda = 0$ }

Component  $ V_{0 +}^a  (k )$ is not gauge invariant but from its gauge transformation  (\ref{eq:GtrV}) we can conclude that scalar   $\varphi_1 = k_a V_{0 +}^a  (k )$ is gauge invariant.
Component  $ V_{0 -}^a  (k )$ is already  gauge invariant  and unique scalar which we can construct from this vector is  $\varphi_2 = q_a V_{0 -}^a  (k )$.
These two  scalars  are first two invariants in  (\ref{eq:Scgi}).  Since they are  gauge invariant   there are no  restrictions  on scalar fields $\varphi_1$  and $\varphi_2$ and
we can use them  as physical fields.  These two fields fill  two dimensional eigenspace with $ \Xi_0$ and they have  together  two degrees of freedom.
They satisfy massless    Klein-Gordon equation.

\subsection{Action  for  massless vector  field with helicities  $\lambda = \pm 1$ produces  Maxwell equations }

The electromagnetic interaction is symmetric under space inversion.  Therefore,  instead of two components $ V_{+ 1}^a  (k)$  and $ V_{- 1}^a  (k )$  which are incomplete irreducible representation  of  proper  Poincare  group  we will introduce one vector field $A^a (k)$ which  is incomplete irreducible representation of  wider group which includes space inversion.
Using (\ref{eq:GSbh})  we obtain
\begin{eqnarray}\label{eq:GSbh1v}
 A^a   (k ) =   \alpha  V^a_{+ 1}  (k )  +  \beta  V^a_{- 1}  (k )  \, .  \qquad
\end{eqnarray}
We also  introduce one parameter   $\Omega (k ) =  \alpha \Omega_+ (k )  +  \beta \Omega_- (k ) $  instead of two components  $\Omega_+ (k )$  and  $\Omega_- (k )$   related by space inversion.

As we have already seen representations $ V_{+ 1}^a  (k)$  and $ V_{- 1}^a  (k )$ are one dimensional which means that  each of them has one degree of freedom. So,  field $A^a (k)$
has two degrees of freedom  and  as we will see  it describes the  photon.

Now, we are ready to introduce action  as gauge invariant expressions of $ A^a  (k )$.  First we  perform  transition from standard momentum $k^a $ to arbitrary one $p^a$
and then to coordinate  dependent   fields  using $p^a = i \partial^a$.  For real field $A^a (x)$, according to   (\ref{eq:Ref1})  and  (\ref{eq:GSbh1v})  we have
\begin{eqnarray}\label{eq:}
A^a (x) =  \int d^4 p  \Big( A^a (p)  +  A^a (- p)   \Big) e^{- i p x }   \nonumber \\
=  \int d^4 p  \Big[   \alpha  \Big(   V^a_{+ 1}  (p )  +   V^a_{+ 1}  (- p )   \Big)  +   \beta \Big(   V^a_{- 1}  (p )   +   V^a_{- 1}  (- p )      \Big)     \Big] e^{- i p x }     \, .
\end{eqnarray}
With the help of  second line in  (\ref{eq:GtrV})  we obtain
\begin{eqnarray}\label{eq:}
\delta  A^a (x)
=  \int d^4 p \,   p^a  \Big[   \alpha   \Big(  \Omega_{+ 1}  (p )  -   \Omega_{+ 1}  (- p )   \Big)
+   \beta \Big(   \Omega_{- 1}  (p )   -  \Omega_{- 1}  (- p )      \Big)     \Big] e^{- i p x }             \, ,
\end{eqnarray}
and using  (\ref{eq:Om1cr})  and    (\ref{eq:om1x})
\begin{eqnarray}\label{eq:GSbh1}
\delta  A^a (x)
=  \partial^a     \Big(  \alpha  \Omega_{+ 1} ( x )  +  \beta \Omega_{- 1}  (x )   \Big)  =  \partial^a  \Omega (x)       \, .
\end{eqnarray}

Action  should depend only on the gauge invariant combination of vector fields, in fact of field strength
$ F_{a b} = \partial_a A_b - \partial_b A_a $.   Since it   must be scalar we can take
\begin{eqnarray}\label{eq:AcI0}
I_0 =  - \frac{1}{4} \int d^4 x    F_{a b} F^{a b}   \, .
\end{eqnarray}

It is not obvious  that field equation, $\partial^2 A^a - \partial^a  \partial_b A^b = 0 $, which follows from    (\ref{eq:AcI0})  has two degrees of freedom. The gauge transformation
 (\ref{eq:GSbh1})  has one parameter. We can use  Lorentz  gauge fixing $\partial_a A^a = 0$  which kills one component of vector field  and after that
equation of motion takes the  form  $\partial^2 A^a = 0 $. But now both gauge fixing and field equation are invariant under additional special gauge transformation  $\delta A^a = \partial^a \Omega_0$,
such that  $\partial^2 \Omega_0 = 0$.  Corresponding gauge fixing kills  additional degree of freedom and consequently electromagnetic field $A^a (x)$ has two degrees of freedom.

Interaction with  other fields  can be described  by the action
\begin{eqnarray}\label{eq:Inm}
I_{int} \Big( A^a \Big)  =  \int d^4 x   A^a J_a  \, ,
\end{eqnarray}
with requirement that it is gauge invariant   $ I_{int} \Big( A^a + \delta  A^a  \Big) =  I_{int} \Big( A^a  \Big)$.
After partial integration we obtain  $ \int d^4 x    \Omega  \,   \partial^a   J_a     =     0  $  and for arbitrary $\Omega$  it produces condition on current to be conserved  $\partial^a J_a  = 0$.

The complete action is
\begin{eqnarray}\label{eq:}
I  =  I_0  +  I_{int} =  \int d^4 x  \Big(   - \frac{1}{4}   F_{a b} F^{a b}   +   A^a J_a   \Big)    \, .
\end{eqnarray}
Variation with respect to $A^a$  produces Maxwell equations.

\section{Second rank tensor  }
\setcounter{equation}{0}

In this section we will consider general features of second rank tensor.

\subsection{Projection operators}

In the  case  of  second rank tensor  we have $A \to (a b ) $ and $\Psi^A \to T^{a b }$.
We will keep notations  $(P_i)^{a b}{}_{c d}$ for projectors and  $(S_{1 2})^{a b}{}_{c d}$ for spin operator of  second ranktensor. Since we will use  abbreviated notation
we introduce new notations for  vector operators,  $( P_{ 0} )^a{}_b  = (\pi_{ 0} )^a{}_b $  and $( P_{ \pm 1} )^a{}_b  =  (\pi_\pm )^a{}_b $ for projectors  and
 $  (S_{1 2})^a{}_b = \rho^a{}_b$ for spin operator.

According  to  (\ref{eq:BEder})  in abbreviated notation  representation of spin operator for  second rank tensor  takes a form
\begin{eqnarray}\label{eq:calSn1c}
( S_{1 2})^{a b }{}_{c d}
=   \rho^a{}_c      \delta^b_d +  \delta^a_c      \rho^b{}_d
=   \Big( \rho   \delta  +  \delta      \rho  \Big)^{a b }{}_{c d}       \,  ,     \qquad
\end{eqnarray}
 where $\rho^a{}_b $  has been defined in   (\ref{eq:SOvf}).
Note that multiplication in abbreviated notation means that we multiply first term with first term and  second term with second term
$  (A \ast B )  (C \ast D ) =  (A C \ast B D) $.

Using relations $\rho = \pi_+ - \pi_- $   and  $\delta =  \pi_0 + \pi_+ + \pi_- $   we  obtain (for simplicity we will omit indices)
\begin{eqnarray}\label{eq:S12D}
 S_{1 2}  = P_1 + 2 P_2 -  P_{-1} - 2 P_{-2}    \,   ,    \qquad  \delta \delta  =    P_1 +  P_2 + P_{-1} + P_{-2}  +  P_0  \,  ,  \qquad
\end{eqnarray}
where
\begin{eqnarray}\label{eq:PO02}
P_0 =  \pi_0    \pi_0   +   \pi_+  \pi_-  +  \pi_-  \pi_+     \, ,  \qquad      P_{\pm 1} =    \pi_0 \pi_\pm  + \pi_\pm  \pi_0 \, , \qquad
     P_{\pm 2} =   \pi_\pm  \pi_\pm       \,   .    \qquad
\end{eqnarray}

Using the fact that $\pi_0,  \pi_+$  and $ \pi_- $ are projectors we can conclude that
 $( P_i )^{a b}{}_{c d}$    $(i = 0, \pm 1, \pm2 )$  are projectors  also.

\subsection{Solutions of   $S_{1 2} $ equation   }

For second rank tensor  characteristic matrix takes the form
\begin{eqnarray}\label{eq:}
 S_{1 2} -  \lambda  \sigma \delta \delta   =  - \lambda \sigma P_0  + (1 - \lambda \sigma )  P_1 +  (2 - \lambda \sigma ) P_2  -   (1 + \lambda \sigma ) P_{-1}   -  (2 + \lambda  \sigma)  P_{-2}    \,   .    \qquad
\end{eqnarray}
Since  $( P_i )^{a b}{}_{c d}$   are projectors,    characteristic  polynomial  is already factored  and we have
\begin{eqnarray}\label{eq:Smladesr}
\det \Big(  S_{1 2} - \lambda \sigma \delta \delta \Big)
=   \Big( - \lambda \sigma \Big)^{d_0}    \Big( 1 - \lambda \sigma  \Big)^{d_1}  \Big( 1 + \lambda \sigma \Big)^{d_{- 1}}  \Big( 2 - \lambda  \sigma \Big)^{d_2}   \Big( 2 + \lambda \sigma \Big)^{d_{- 2}}
\prod_{i = - 2}^2  \det  P_i    = 0        \,   ,    \qquad  \qquad
\end{eqnarray}
where $d_i$ are  dimensions of eigenspaces  $\Xi_i$.
Dimension of subspace projected by $\pi_0$ is $2$ while  dimensions of   subspaces   projected by    $\pi_+$   and   $\pi_-$  are $1$, so that  in our case
$ d_0 = 6 \, , \,   d_{\pm 1} = 4 \, ,  \,    d_{\pm  2} = 1 $. It is in agreement with general relation since $6 + 2 ( 4 + 1) =  4^2$.

Then with the help of relation    $\sigma^2 = 1$  we obtain
\begin{eqnarray}\label{eq:Smladed}
\det  \Big(  S_{1 2} - \lambda \sigma \delta \delta \Big)
=   \Big(- \lambda \Big)^6     \Big( 1 - \lambda^2   \Big)^4   \Big(4 - \lambda^2  \Big)  \prod_{i = - 2}^2  \det  P_i      = 0        \,   .    \qquad
\end{eqnarray}
Using  the fact that  $\det P_i \ne 0$ we can find   that spectrum of  helicities is
\begin{eqnarray}\label{eq:}
\lambda_0 = 0 \, , \qquad    \lambda_{\pm 1} = \pm 1 \, , \qquad   \lambda_{\pm 2} =  \pm 2    \, . \qquad
\end{eqnarray}

Consequently, the eigenfunctions   of   operator  $S_{1 2} $  in full notation   have the  form
\begin{eqnarray}\label{eq:}
T_i^{a b}  (k) =  (P_i)^{a b }{}_{c d}   T^{c d}  (k)  \, ,  \qquad    (i = 0, \pm 1, \pm 2  )
\end{eqnarray}
where  projection operators  $(P_i)^{a b }{}_{c d}$ are defined in (\ref{eq:PO02}).

\section{Second rank tensor with  helicities  $\lambda = \pm 2$   and   general relativity in weak field approximation  }
\setcounter{equation}{0}

Second rank tensor  with  helicities  $\lambda = \pm 2$ is  of particular importance  since its  symmetric  part (metric)  contribute to  general relativity.

\subsection{Gauge  transformation for highest   helicities $\lambda = \pm 2$ of  second rank tensors  }

Each component of the second rank tensor  with highest helicities    $\lambda = \pm 2 \,$
\begin{eqnarray}\label{eq:Hhpm2}
T_{\pm 2}^{a b} (k) =   (P_{\pm 2})^{a b }{}_{c d}   T^{c d} (k) =     ( \pi_{ \pm 1} )^a{}_c   ( \pi_{ \pm 1} )^b{}_d  T^{c d}  (k)  =      \frac{1}{4}   (\breve{p}_\mp)_c   (\breve{p}_\mp)_d    T^{c d} (k)  \breve{p}_\pm^a      \breve{p}_\pm^b       \, ,  \qquad
\end{eqnarray}
is  one dimensional representation and consequently it has one degree of freedom. Note that by definition it is symmetric tensor  and relation  (\ref{eq:Hln})  is satisfy.

With the help of   (\ref{eq:BEder})   we can express  little group generators for massless  second rank tensors  $ (\Pi_1 )^{a b }{}_{c d }$  and  $ (\Pi_2 )^{a b }{}_{c d }$
in terms of corresponding  generators   for vectors  $ (\Pi_1)^a{}_b$ and $ (\Pi_2)^a{}_b$
\begin{eqnarray}\label{eq:}
  (\Pi_1 )^{a b }{}_{c d } =   (\Pi_1)^a{}_c \delta^b_d     +  \delta^a_c  (\Pi_1)^b{}_d         \, ,  \qquad
 (\Pi_2 )^{a b }{}_{c d } =   (\Pi_2)^a{}_c \delta^b_d     +  \delta^a_c  (\Pi_2)^b{}_d          \, .
\end{eqnarray}
In that case instead  of general expression  for  gauge transformation  (\ref{eq:Txyvg})  we have
\begin{eqnarray}\label{eq:Txyvgs}
\delta  T_i^{a b}   ( \varepsilon_1, \varepsilon_2 )  (k)
 =  i \Big(  \varepsilon_1 \Pi_1  +  \varepsilon_2 \Pi_2  \Big)^{a b}{}_{c d}  \,    T_i^{c d}  (k)   \, .   \qquad
\end{eqnarray}
Using expressions for  highes helicities   (\ref{eq:Hhpm2})  and gauge transformation for basic vector   $ \delta \breve{p}_\pm^a  =   \varepsilon_\pm k^a $     we obtain
\begin{eqnarray}\label{eq:GtrT2}
\delta   T_{\pm 2}^{a b}  (k)  =   k^a  \Omega_\pm^b +   k^b \Omega_\pm^a                \, ,  \qquad
  \Omega_\pm^a   (k) =     \frac{1}{4}   (\breve{p}_\mp)_c   (\breve{p}_\mp)_d    T^{c d}   \breve{p}_\pm^a  \varepsilon_\pm          \, .  \qquad
\end{eqnarray}
Note that again ${\cal P} T_{\pm 2}^{a b}  (k) = T_{\mp 2}^{a b}  (k)$  and ${\cal P}  \Omega_\pm^a = \Omega_\mp^a $.

Here we came to a similar conclusion for  highest helicities of second rank tensor  as in the case of vector  field: it is not physical since it is not Lorentz invariant.

\subsection{General relativity in weak field approximation   }

As in the case of vector fields, components of  massless second rank tensors  of opposite helicities $\lambda =  2$ and   $\lambda = - 2$   are related by transformation of space inversion.
Since   gravitation interaction  is symmetric under space inversion   it is common to treat   both components of incomplete irreducible representation  of  proper  Poincare  group
as a single particle called   graviton.
So, instead of two  polarizations  $ T_{+ 2}^{a b}  (k )$  and $ T_{- 2}^{a b}  (k )$, both with one degree of freedom,   we will introduce one  symmetric tensor
$h^{b a} (k) = h^{a b} (k) $  with two   degrees of freedom, which  is incomplete irreducible representation of Poincare group with  space inversion.
Using (\ref{eq:GSbh})  we obtain
\begin{eqnarray}\label{eq:GSbh10}
 h^{a b}  (k ) =   \alpha  T^{a b}_{+ 2}  (k )  +  \beta  T^{a b}_{- 2}  (k )  \, .  \qquad
\end{eqnarray}
We also  introduce one parameter   $\Omega^a (k ) =  \alpha \Omega_+^a  (k ) +  \beta \Omega_-^a (k )$  instead of two components  $\Omega_+^a (k )$  and  $\Omega_-^a (k )$   related by space inversion.

If we go from standard momentum frame  to arbitrary frame  and then to coordinate  dependent   fields,  $k^a  \to p^a \to    i \partial^a$,
we  can obtain  gauge transformation in coordinate representation. For real field,   according to   (\ref{eq:Ref1})  and  (\ref{eq:GSbh10})     we have
\begin{eqnarray}\label{eq:}
h^{a b} (x) =  \int d^4 p  \Big( h^{a b} (p)  +  h^{a b} (- p)   \Big) e^{- i p x }   \nonumber \\
=  \int d^4 p  \Big[   \alpha  \Big(   T^{a b}_{+ 2}  (p )   +   T^{a b}_{+ 2}    (- p )   \Big)  +   \beta \Big(   T^{a b}_{- 2}   (p )   +   T^{a b}_{- 2}   (- p )      \Big)     \Big] e^{- i p x }     \, .
\end{eqnarray}
With the help of    (\ref{eq:GtrT2})  we obtain gauge transformation
\begin{eqnarray}\label{eq:}
\delta  h^{a b} (x)
=  \int d^4 p \, \Big\{  p^a  \Big[   \alpha   \Big(  \Omega_{+ }^b  (p )  -   \Omega_{+}^b  (- p )   \Big)
+   \beta \Big(   \Omega_{- }^b  (p )   -  \Omega_{- }^b  (- p )      \Big)     \Big]      \nonumber \\
 + p^b  \Big[   \alpha   \Big(  \Omega_{+ }^a  (p )  -   \Omega_{+ }^a  (- p )   \Big)
+   \beta \Big(   \Omega_{- }^a  (p )   -  \Omega_{- }^a  (- p )      \Big)     \Big]           \Big\}   e^{- i p x }             \, ,
\end{eqnarray}
and using  (\ref{eq:Om1cr})  and    (\ref{eq:om1x})
\begin{eqnarray}\label{eq:Gthab1}
\delta  h^{a b} (x)
=  \partial^a     \Big(  \alpha  \Omega_{+ }^b ( x )  +  \beta \Omega_{- }^b  (x )   \Big)   +     \partial^b     \Big(  \alpha  \Omega_{+ }^a ( x )  +  \beta \Omega_{- }^a  (x )   \Big)
=  \partial^a  \Omega^b (x)   + \partial^b  \Omega^a (x)     \, .  \qquad
\end{eqnarray}

In order to get field equation we need  gauge invariant  function of  tensor $ h^{a b} (x)$.
It is linear combination of massless Klein-Gordon equation and supplementary conditions $ G^b $
\begin{eqnarray}\label{eq:}
\partial^2 h_{a b}  = 0  \, ,  \qquad  G^b = \partial_a h^{a b} -  \frac{1}{2} \partial^b h^a{}_a =  0  \, .
\end{eqnarray}
The  supplementary conditions are in fact gauge fixing which  kill all unwanted degrees of freedom.

 These equations can be combined  into one equation as linear combination of   Klein-Gordon equation  and supplementary conditions
\begin{eqnarray}\label{eq:}
\partial^2 h_{a b}  +  \alpha   \Big( \partial_a G_b +  \partial_b G_a      \Big)   =  0  \, .
\end{eqnarray}

It must be gauge invariant and using gauge transformation of gauge fixing condition $\delta G_a = \partial^2  \Omega_a $  we obtain   $\alpha  = - 1$.

Consequently,  equation of motion takes a form  of  Ricci curvature
\begin{eqnarray}\label{eq:}
R_{a b}  =  \partial^2 h_{a b} -   \Big( \partial_a G_b +  \partial_b G_a      \Big)
 =  \partial^2 h_{a b} -   \partial_a  \partial^c h_{c b}  -   \partial_b  \partial^c h_{c a} +  \partial_a  \partial_b  h^c{}_c   =  0  \, .
\end{eqnarray}
Then scalar curvature is
\begin{eqnarray}\label{eq:}
R  =  \eta^{a } R_{a b}  = \partial^2 h - 2  \partial^a G_a
=  2 \Big(    \partial^2 h -   \partial^a  \partial^b h_{a b}   \Big)    \, ,
\end{eqnarray}
and  Einstein tensor for free field equation
\begin{eqnarray}\label{eq:FeET}
G^{a b} (x) = R^{a b} - \frac{1}{2}  \eta^{a b} R = \partial^2 \Big(  h_{a b} -  \frac{1}{2} \eta^{a b} h  \Big)  -   \Big( \partial_a G_b +  \partial_b G_a  \Big)  + \eta^{a b}  \partial^c G_c       \nonumber \\
=  \partial_c   \partial^b   h^{a c}   - \partial^2  h^{a b}   -   \partial^a   \partial^b   h +  \partial^a   \partial_c   h^{c b}
- \eta^{a b}   \Big(  \partial_c   \partial_d   h^{c d}   - \partial^2  h \Big) =  0   \, .  \qquad
\end{eqnarray}
All  these tensors  can be obtained as small  perturbation  around flat metric using   relation  $g_{a b} =  \eta_{a b} + h_{a b}$,  see Refs.\cite{LL, WG}.

Field equation   (\ref{eq:FeET})  can be obtained from the action
\begin{eqnarray}\label{eq:}
 I_0  (h^{a b}) =    \int  d^4 x    \Big( \frac{1}{2}  \partial_c  h_{a b}  \partial^c  h^{a b}   -   \partial_c  h_{a b}  \partial^b  h^{a c}
+  \partial_c  h^{a c}  \partial_a h - \frac{1}{2}    \partial_a h \partial^a h    \Big)     \, .
\end{eqnarray}

Similarly as in the case of vector field it can be shown  (see for example Ref.\cite{WG})   that  field $h^{a b} $ which satisfy  this equation has two degrees of freedom.
The symmetric tensor  $h^{a b} $  has $10$ components.
To fix    gauge invariance   (\ref{eq:Gthab1})  we can use  $4$  gauge fixings conditions  $ G^b = \partial_a h^{a b} - \frac{1}{2} \partial^b h^a{}_a = 0 $  which kills $4$  component of tensor  field  and after that
equation of motion takes the form  $\partial^2 \Big(  h^{a b}  -\frac{1}{2}  \eta^{a b} h \Big)  = 0 $.
 But now both gauge fixing and field equation are invariant under additional special gauge transformation  $\delta  h^{a  b}  (x) =   \partial^a  \Omega_0^b (x)  +   \partial^b  \Omega_0^a  (x) $,
such that  $\partial^2 \Omega_0^a = 0$.  Corresponding gauge fixing kills  additional  $4$ degree of freedom and consequently gravitation  field $h^{a b} (x)$ has two degrees of freedom.

Interaction with matter  field $\Phi^A$   we can describe with the action quadratic in $h^{a b}$
\begin{eqnarray}\label{eq:Inmhab}
 I_{int}  ( h^{a b} ) =  \int d^4 x  {\cal L} ( h^{a b}, \Phi^A )   \, .
\end{eqnarray}
Note that the integration measure is  the flat Minkowski measure $d^4 x$, as Lagrangian is already quadratic in  $h^{a b}$.

From  requirement that $I_{int} ( h^{a b} )$   is gauge  invariant  $ I_{int}  ( h^{a b} + \delta  h^{a b})  =    I_{int}  ( h^{a b} )  $  we have
\begin{eqnarray}\label{eq:}
  \int  d^4 x  \delta h^{a b}  \frac{ \partial  {\cal L}  }{\partial  h^{a b}  }  =   \int  d^4 x  \Big(  \partial^a  \Omega^b   +   \partial^b  \Omega^a    \Big)  \Theta_{a b}  = 0   \, .  \qquad
\Theta_{a b}  =  \frac{ \partial  {\cal L}  }{\partial  h^{a b}  }
\end{eqnarray}
Using partial integration and  fact that  $\Theta_{a b}$  is symmetric tensor   we obtain   that   $\Theta_{a b}$  is the conserved energy-momentum tensor,   $\partial_a  \Theta^{a b} = 0$.

The complete action is $I (h^{a b}) = I_0  (h^{a b}) + I_{int} (h^{a b})$. Its variation with respect to  $h^{a b}$  produces complete Einstein  equations  in weak field approximation
\begin{eqnarray}\label{eq:FEeq12p}
G_{a b}  (h^{a b}) =    \Theta_{a b}   \, .
\end{eqnarray}

\section{Second rank tensor with  helicities   $\lambda = \pm 1$  produces Maxwell equations}
\setcounter{equation}{0}

Beside highes helicities   $\lambda = \pm 2$,  second rank tensor  has also components with  helicities   $\lambda = \pm 1$. We expect that these components describe  Maxwell equations in terms of field strength.

\subsection{Second rank tensor with  helicities  $\lambda = \pm 1$  }

The components   of second rank tensor  with  helicities $\lambda  = \pm 1$  have a form
\begin{eqnarray}\label{eq:}
T_{\pm 1}^{a b}  (k)  =  (P_{\pm 1})^{a b }{}_{c d}   T^{c d} (k) =   \Big(  (\pi_0)^a{}_c    (\pi_{\pm })^b{}_d  +   (\pi_{\pm })^a{}_c    (\pi_0)^b{}_d  \Big)   T^{c d}    (k)              \, .
\end{eqnarray}
Since $\pi_0$ project to two dimensional space and   $\pi_+$ and $\pi_-$   to one dimensional space,  both components $T_{\pm 1}^{a b}  (k)  $  are four dimensional representations and they  have four degrees of freedom.

To avoid influence of  nontrivial  gauge transformations of $( \pi_{ 0 +} )^a{}_b$   we will save  only $( \pi_{ 0 -} )^a{}_b$ part  (gauge  invariant     part of
$( \pi_{ 0 } )^a{}_b$)   which means  that we will take  $( \pi_{ 0 } )^a{}_b  \to ( \pi_{ 0 -} )^a{}_b$.  With that choice  we obtain
\begin{eqnarray}\label{eq:}
T_{\pm 1 (-)}^{a b} (k) =  (P_{\pm 1 (-)})^{a b }{}_{c d}   T^{c d} (k)  =   \Big(  (\pi_{0 -})^a{}_c    (\pi_{\pm })^b{}_d  +   (\pi_{\pm })^a{}_c    (\pi_{0 -})^b{}_d  \Big)   T^{c d}  (k)                \, .
\end{eqnarray}
Note that now each component   $T_{+ 1 (-)}^{a b} (k)$  and  $T_{- 1 (-)}^{a b} (k)$ has  two  degrees of freedom.

Using expressions for projectors  we get
\begin{eqnarray}\label{eq:}
T_{\pm 1 (-)}^{a b} (k) =  - \frac{1}{4}   \Big(  q_c  (\breve{p}_\mp)_d   T^{c d} (k)  \Big)   k^a  \breve{p}_\pm^b  - \frac{1}{4}   \Big(  (\breve{p}_\mp)_c    q_d  T^{c d} (k) \Big)   \breve{p}_\pm^a    k^b   \, .
\end{eqnarray}
Since $k_a k^a = 0$  and  $k_a  \breve{p}_\pm^a = 0$  there are  constraints  $ k_a T_{\pm 1 (-)}^{a b} (k) = 0 $.
The fields  $T_{\pm 1 (-)}^{a b} (k) $  with helicities $\lambda = \pm 1$  contain   one basic vector $\breve{p}_\pm^a$ and
consequently  relation (\ref{eq:Hln})  is satisfy.

For symmetric tensor we obtain
\begin{eqnarray}\label{eq:STpm1}
(T^S_{\pm 1 (-)})^{(a b)} (k) =   - \frac{1}{4}   \Big(  q_c   (\breve{p}_\mp)_d   T^{(c d)}  (k) \Big) \Big(   k^a  \breve{p}_\pm^b   +   k^b  \breve{p}_\pm^a  \Big)
\equiv   \mathbf{T}_{\mp 2}^S  \Big(   k^a  \breve{p}_\pm^b   +   k^b  \breve{p}_\pm^a  \Big)      \, ,   \qquad
\end{eqnarray}
or
\begin{eqnarray}\label{eq:STpm1r}
(T^S_{\pm 1 (-)})^{(a b)} (k) =     k^a  \varphi_\pm^b (k)  +   k^b  \varphi_\pm^a   (k)     \, ,   \qquad
\varphi_\pm^a =   \mathbf{T}_{\mp 2}^S  \breve{p}_\pm^a  \, ,
\end{eqnarray}
and for antisymmetric one
\begin{eqnarray}\label{eq:Asljj}
(T^A_{\pm 1 (-)})^{[a b]}  (k) =      - \frac{1}{4}   \Big(  q_c   (\breve{p}_\mp)_d   T^{[c d]} (k)  \Big) \Big(   k^a  \breve{p}_\pm^b  -   k^b  \breve{p}_\pm^a  \Big)
\equiv    \mathbf{T}_{\mp 2}^A  \Big(   k^a  \breve{p}_\pm^b   -   k^b  \breve{p}_\pm^a  \Big)       \, ,   \qquad
\end{eqnarray}
or
\begin{eqnarray}\label{eq:Asljjr}
(T^A_{\pm 1 (-)})^{[a b]}  (k) =       k^a  A_\pm^b  (k) -   k^b  A_\pm^a   (k)      \,  ,  \qquad
A_\pm^a  =     \mathbf{T}_{\mp 2}^A  \breve{p}_\pm^a  \, .
\end{eqnarray}
Here $\varphi_\pm^a$  and  $A_\pm^a $  are new fields defined above. Note that both $\varphi_\pm^a$  and  $A_\pm^a $ are proportional to $\breve{p}_\pm^a$, as well as fields $V_\pm^a$ in
Eq.(\ref{eq:Vfbv}).  According to scalar product of basic vectors $k_a \breve{p}_\pm^a  = 0$ we can conclude that  $k_a \varphi_\pm^a = 0 $   and  $k_a A_\pm^a = 0 $, which means that these solutions are
in Lorentz gauge. We will see that  these  fields  describe electromagnetic interaction.

Finally,  each component   $(T^S)_{+ 1 (-)}^{a b} (k)$,   $(T^S)_{- 1 (-)}^{a b} (k)$,   $(T^A)_{+ 1 (-)}^{a b} (k)$ and    $(T^A)_{- 1 (-)}^{a b} (k)$ have one  degree of freedom.

\subsection{Gauge transformation for second rank tensor  with  helicities  $\lambda = \pm 1$ and Maxwell equations }

Using gauge transformations for basic vectors $\delta k^a = 0$ and  $\delta  \breve{p}_\pm^a =   k^a \varepsilon_\pm $ we can obtain  gauge transformations for symmetric and antisymmetric parts of
second rank tensor with  helicities $\lambda  = \pm 1$
\begin{eqnarray}\label{eq:Gtrst}
\delta  (T^S_{\pm 1 (-)})^{(a b)} (k) =k^a k^b  \omega_\pm  \, , \qquad  \omega_\pm (k) = - \frac{1}{2}   \Big(  q_c  (\breve{p}_\mp)_d   T^{(c d)} (k)  \Big)   \varepsilon_\pm      \, ,  \qquad \nonumber \\
\delta  (T^A_{\pm 1 (-)})^{[a b]}  (k) =  0   \, . \qquad  \qquad  \qquad  \qquad
\end{eqnarray}
So, symmetric tensor is gauge dependent  and antisymmetric one is gauge  invariant.

\subsubsection{Symmetric tensor with helicities  $\lambda = \pm 1$}

Note that under space inversion ${\cal P} (T^S_{\pm 1 (-)})^{(a b)} (k) =  (T^S_{\mp 1 (-)})^{(a b)} (k)$,  ${\cal P}  \varphi^a_\pm (k) = \varphi^a_\mp (k) $   and
${\cal P}  \omega_\pm  (k) = \omega_\mp (k) $.  So,   according to  (\ref{eq:GSbh})   instead of two components $ (T^S_{ + 1 (-)})^{(a b)}  (k )$  and $(T^S_{-1 (-)})^{(a b)}  (k )$
we will introduce one  symmetric tensor
\begin{eqnarray}\label{eq:GSv12}
 \varphi^{a b}  (k ) =   \alpha  (T^S_{ + 1 (-)})^{(a b)}  (k )   +  \beta  (T^S_{-1 (-)})^{(a b)}  (k )  \, ,  \qquad
\end{eqnarray}
and  instead of two components   $\varphi^a_\pm$   we will introduce one vector  field
\begin{eqnarray}\label{eq:GSv11}
 \varphi^a  (k ) =   \alpha  \varphi^a_+   (k )   +  \beta  \varphi^a_-   (k )  \, ,  \qquad
\end{eqnarray}
which  are incomplete  irreducible representation of Poincare group with  space inversion.
The representation of the field  $\varphi^{a b} (k) $ as well as of  $\varphi^a  (k )$ is two dimensional and consequently they have  two degrees of freedom.
Since the parameters $ \omega_\pm (k) $ are also connected by parity transformation we can introduce one parameter $\omega (k) = \alpha \omega_+  (k) +    \beta \omega_- (k)$.

If we go to arbitrary frame  and coordinate  dependent   fields,   $k^a  \to p^a \to    i \partial^a$,  we obtain  expression  for symmetric tensor
$ \varphi^{a b} (x)$ in terms of vector field  $\varphi^a (x)$ and theirs gauge transformations.    For real field,   according to   (\ref{eq:Ref1})  and  (\ref{eq:GSv12})     we have
\begin{eqnarray}\label{eq:vpabf}
\varphi^{a b} (x) =  \int d^4 p  \Big( \varphi^{a b} (p)  +  \varphi^{a b} (- p)   \Big) e^{- i p x }   \nonumber \\
=  \int d^4 p  \Big[   \alpha  \Big(   (T^S_{ + 1 (-)})^{(a b)}    (p )   +   (T^S_{+1 (-)})^{(a b)}    (- p )   \Big)
+   \beta \Big(   (T^S_{-1 (-)})^{(a b)}    (p )   +  (T^S_{-1 (-)})^{(a b)}   (- p )      \Big)     \Big] e^{- i p x }     \, .  \qquad
\end{eqnarray}
Then with the help of  (\ref{eq:STpm1r})  we obtain
\begin{eqnarray}\label{eq:}
 \varphi^{a b} (x)
=  \int d^4 p \, \Big\{  p^a  \Big[   \alpha   \Big(  \varphi_{+ }^b  (p )  -   \varphi_{+}^b  (- p )   \Big)
+   \beta \Big(   \varphi_{- }^b  (p )   -  \varphi_{- }^b  (- p )      \Big)     \Big]      \nonumber \\
 + p^b  \Big[   \alpha   \Big(  \varphi_{+ }^a  (p )  -   \varphi_{+ }^a  (- p )   \Big)
+   \beta \Big(   \varphi_{- }^a  (p )   -  \varphi_{- }^a  (- p )      \Big)     \Big]           \Big\}   e^{- i p x }             \, ,
\end{eqnarray}
and using  (\ref{eq:Om1cr})  and    (\ref{eq:om1x})
\begin{eqnarray}\label{eq:Gthab}
\varphi^{a b} (x)
=  \partial^a     \Big(  \alpha  \varphi_{+ }^b ( x )  +  \beta \varphi_{- }^b  (x )   \Big)   +     \partial^b     \Big(  \alpha  \varphi_{+ }^a ( x )  +  \beta \varphi_{- }^a  (x )   \Big)
=  \partial^a  \varphi^b (x)   + \partial^b  \varphi^a (x)     \, .  \qquad
\end{eqnarray}

To obtain gauge transformation we use    (\ref{eq:Gtrst}) and  (\ref{eq:vpabf})
\begin{eqnarray}\label{eq:}
\delta  \varphi^{a b} (x)  (x)
=  \int d^4 p \,  p^a p^b \Big[   \alpha   \Big(  \omega_{+ } (p )  -   \omega_{+}  (- p )   \Big)
+   \beta \Big(   \omega_{- }  (p )   -  \omega_{- }  (- p )      \Big)     \Big]     e^{- i p x }             \, ,
\end{eqnarray}
and  with the help of   (\ref{eq:Om2cr})  and    (\ref{eq:om1x})
\begin{eqnarray}\label{eq:Gthabd}
\delta  \varphi^{a b} (x)
=  \partial^a  \partial^b   \Big(  \alpha  \omega_{+ } ( x )  +  \beta \omega_{- }  (x )   \Big)   =  \partial^a  \partial^b  \omega (x)      \, .  \qquad
\end{eqnarray}

So, gauge transformation of field $ \varphi^a (x)$ takes the  form
\begin{eqnarray}\label{eq:}
\delta  \varphi^a  (x) =   \partial^a   \omega_1  (x)   \, ,  \qquad   \omega_1 (x) = \frac{1}{2} \omega   (x) \, .
\end{eqnarray}

The field $\varphi^a  (x)$  behaves like  electromagnetic field. It is vector field, it has the same  helicity, the same  gauge transformation  and the same number of degrees of freedom.
So, this  field  should not be considered as new field.

\subsubsection{Antisymmetric tensor with   helicities  $\lambda = \pm 1$   and   Maxwell equations}

Antisymmetric part is gauge invariant and consequently Lorentz invariant and we can use it as physical field.
Instead of two components $ (T^A_{ + 1 (-)})^{[a b]}  (k )$  and $(T^A_{-1 (-)})^{[a b]}  (k )$,  according to   (\ref{eq:GSbh})    we will introduce one  antisymmetric tensor
\begin{eqnarray}\label{eq:GSv12A}
F^{a b}  (k ) =   \alpha  (T^A_{ + 1 (-)})^{[a b]}  (k )   +  \beta  (T^A_{-1 (-)})^{[a b]}  (k )  \, ,  \qquad
\end{eqnarray}
which  is irreducible representation of Poincare group with  space inversion.
Similarly,  from  (\ref{eq:Asljjr})  we have
\begin{eqnarray}\label{eq:Fabk}
F^{a b}  (k )  =       k^a  A^b  (k) -   k^b  A^a   (k)      \,  ,  \qquad
A^a  (k) =  \alpha  A_+^a  (k)  + \beta A_-^a   (k)      \, .
\end{eqnarray}

Using    (\ref{eq:Asljj}) and scalar product of basic vectors   we can conclude that   there  are two constraints on field $F^{a b} (k)$
\begin{eqnarray}\label{eq:}
k_a   F^{a b}  (k)  = 0     \,  ,  \qquad    \varepsilon_{a b c d}   k^b    F^{c d} (k) =  0     \, .   \qquad
\end{eqnarray}

If we go to arbitrary frame  and coordinate  dependent   fields,   $k^a  \to p^a \to    i \partial^a$,   in the same way as before  we get
\begin{eqnarray}\label{eq:}
\partial_a   F^{a b}  (x)  = 0     \,  ,  \qquad    \varepsilon_{a b c d}   \partial^b    F^{c d} (x) =  0     \, ,  \qquad
\end{eqnarray}
which are just vacuum  Maxwell equations.
As well as in the previous case the representation of the field  $F^{a b} (x) $  is two dimensional and consequently it  has two degrees of freedom.
We could have expected that,  since it describes   Maxwell equations.

To express  coupling of  electromagnetic fields with   matter (to a conserved current) we must solve second equation which produces
\begin{eqnarray}\label{eq:FabA}
F_{a b} (x)   = \partial_a A_b (x) -   \partial_b A_a  (x)   \, ,  \qquad
\end{eqnarray}
which   also  follows from   (\ref{eq:Fabk}).

Note that   $A_a (x)$ is gauge field since $ F_{a b} (x)$  is invariant under transformation $\delta A_a (x) =  \partial_a \Omega$.  Then interaction with a matter can be express as in  (\ref{eq:Inm}).

%%%%%

\section{Fourth  rank tensor  with    helicities $\lambda = \pm 2$  produces  general  relativity  in weak field approximation }\label{sec:Frt}
\setcounter{equation}{0}

Since a symmetric second-rank tensor field with helicities  $\lambda = \pm 1$  describes electromagnetic interactions through the field strength tensor, we naturally expect that a fourth-rank tensor field with helicities   $\lambda = \pm 2$ and appropriate symmetry properties should describe gravitational interactions through the Riemann curvature tensor.

For the fourth-rank tensor case, the field index $A$ corresponds to a quadruple index  
  $A \to (a b c d) $  with  $\Psi^A \to T^{a b c d}$. 
In this section, we will maintain the compact notation using index $A$ for simplicity, reverting to the explicit index notation $abcd$ only when discussing specific symmetry properties.

\subsection{Fourth rank tensor  }

We are going to solve equation
\begin{eqnarray}\label{eq:}
\Big( S^A{}_B - \lambda \sigma \delta^A_B  \Big)  T^B  = 0  \, ,
\end{eqnarray}
where $T^A$  is fourth rank tensor,    $T^A = T^{a b c d }$.  The problem comes down to that of  vector fields.
According  to  (\ref{eq:BEder})  in abbreviated notation   spin operator   takes the form
\begin{eqnarray}\label{eq:S12ar}
 (S_{1 2} )^A{}_B
=   \Big( \rho   \delta^3  +  \delta \rho \delta^2    +  \delta^2 \rho \delta + \delta^3 \rho  \Big)^A{}_B    \,   .     \qquad
\end{eqnarray}
Using  expressions  $\rho = \pi_+ - \pi_- $   and  $\delta =  \pi_0 + \pi_+ + \pi_- $  after multiplications we   obtain
\begin{eqnarray}\label{eq:cSrdf}
 (S_{1 2})^A{}_B    =  \sum_{i = - 4}^4   i \Big(   P_i   \Big)^A{}_B  \,   ,     \qquad
\end{eqnarray}
and
\begin{eqnarray}\label{eq:Defg}
\delta^A_B  =  ( \delta^4 )^A{}_B  =  \Big( ( \pi_0 + \pi_+ + \pi_- )^4 \Big)^A{}_B  =  \sum_{i = - 4}^4     \Big( P_i \Big)^A{}_B             \, ,
\end{eqnarray}
where the multiplication factor  $i$  in (\ref{eq:cSrdf})  is consequence of combinatorics.

Here $(P_i)^A{}_B$  is defined as the sum of all possible terms  $\pi_{i_1}   \pi_{i_2}  \pi_{i_3}  \pi_{i_4}  $ such that $\sum_{m=1}^4 i_m = i$  where we assign  $\pi_+  \equiv  \pi_1$  and  
$\pi_-   \equiv \pi_{- 1}$. 
The sequence of multipliers is significant.  For example $\pi_+   \pi_0   \pi_+ \pi_0 $ is different from  $\pi_0   \pi_+   \pi_+ \pi_0 $.   
Since   $\pi_0,  \pi_+$   and  $\pi_- $ are projectors, it follows that all    $P_i$  for $  i = 0,  \pm 1, \pm 2, \pm 3,   \pm 4 $  are also  projectors,  satisfying the orthogonality condition  $P_i P_j   =  \delta_{i j } P_i$.

To obtain spectrum of  helicities we are going to solve consistency condition  that characteristic polynomial vanishes,
$\det ( S_{1 2}   - \lambda \sigma \delta )^A{}_B  =  0    $.
Using   (\ref{eq:cSrdf})  and   (\ref{eq:Defg})    we obtain expression for characteristic polynomial
\begin{eqnarray}\label{eq:Smlaar}
 (S_{1 2})^A{}_B   - \lambda \sigma \delta^A_B
=    \sum_{i = - 4}^4     (i - \lambda \sigma)  ( P_i)^A{}_B              \,   .    \qquad
\end{eqnarray}
Since   $( P_i )^A{}_B$  are  projection  operators,  expression   $S_{1 2} - \lambda \sigma \delta$  is already factored   and we have
\begin{eqnarray}\label{eq:SmladeRt}
\det ( S_{1 2} - \lambda \sigma \delta)
=     \prod_{i = - 4}^4   ( i - \lambda \sigma)^{d_i}   \,    \det  P_i    = 0         \,   ,    \qquad
\end{eqnarray}
where $d_i$ are  dimensions of eigenspaces  $\Xi_i$.  In our case
\begin{eqnarray}\label{eq:}
 d_4  =    1        \, ,   \qquad
 d_3     = 8        \, , \qquad
d_2   =  28   \, , \qquad
 d_1    =   56    \, ,   \qquad  d_0 = 70  \, ,
\end{eqnarray}
with   general relation    $d_i = d_{- i}$.  Note that  $  2 ( 1 + 8 +    28  +    56 )   +   70  =  4^4 $.

Then  with the help of relation $\sigma^2 = 1$  we obtain
\begin{eqnarray}\label{eq:Smlade12}
\det  ( S_{1 2} - \lambda \sigma  \delta)
=   ( - \lambda )^{d_0}   \prod_{i =1}^4   ( i^2 - \lambda^2 )^{d_i}   \,    \prod_{i = -4}^4   \det  P_i   = 0         \,   .    \qquad
\end{eqnarray}
Using the fact that $\det P_i \ne 0$ we can conclude   that
\begin{eqnarray}\label{eq:}
( - \lambda)^{d_0}   \prod_{i =1}^4   ( i^2 - \lambda^2 )^{d_i}     = 0      \,   ,   \qquad
\end{eqnarray}
which produces  spectrum of  helicities for four  rank tensor
\begin{eqnarray}\label{eq:}
\lambda_0 = 0 \, , \qquad     \lambda_{\pm  i} = \pm i     \, .   \qquad  i = (1, 2, 3, 4 )
\end{eqnarray}

The  eigenfunctions, which are our candidates for  irreducible representations,  for four rank massless  tensors  are
\begin{eqnarray}\label{eq:Eif}
T_i^A  (k) =  (P_i)^A{}_B    T^B  (k)  \, .  \qquad    (i = 0, \pm 1, \pm 2 , \pm 3, \pm 4  )
\end{eqnarray}

\subsection{General relativity in terms of  Riemann tensor   }

In this  subsection we will find  irreducible representations of fourth rank tensor with helicity  $\lambda = \pm 2$. First we obtain
irreducible representations of proper Poincare  group,  $T_{\pm 2 (-)}^{a  b c d } (k)  $ and then of wider group which combined  Poincare group and parity,  $ R^{a b c d} (k) $.
The last one  is  linearized  Riemann tensor.  Finally, we obtain   Einstein equations in vacuum.

\subsubsection{Irreducible representations of  proper Poincare  group  }

The fourth rank tensor  with  helicities $\lambda = \pm 2$ has a form
\begin{eqnarray}\label{eq:NHhg}
T_{\pm 2 }^{a b c d} (k)  =    (P_{\pm 2 })^{a b c d }{}_{e f g h }    T^{e f g h }   (k)   \nonumber \\
 =  \Big[  \sum_{\Sigma_1}  (\pi_+)^a{}_e  (\pi_+)^b{}_f  (\pi_0)^c{}_g   (\pi_0)^d{}_h
 +    \sum_{\Sigma_2}     (\pi_+)^a{}_e  (\pi_+)^b{}_f  (\pi_+)^c{}_g   (\pi_-)^d{}_h  \Big]    T^{e f g h }   (k)  \, ,  \qquad
\end{eqnarray}
where  $\Sigma_1$  and  $\Sigma_2$   means sum over  all possible permutations of projectors.

As well as in the case of second rank  tensor with helicities  $\lambda = \pm 1$,  to describe    particles with helicities $\lambda = \pm 2$  with the help of fourth rank  tensor,
in every place  we will  substitute $\pi_0 $ with its   gauge invariant part   $\pi_{0 -}$.  Then   fourth rank  tensor  with  helicities $\lambda = \pm 2$ has two terms: one with   $2$  vectors
$\breve{p}_\pm^a$  and  $2$  vectors  $k^a$  and the other  with   $3$  vectors   $\breve{p}_\pm^a$  and  one  vector   $\breve{p}_\mp^a$
\begin{eqnarray}\label{eq:NHh}
T_{\pm 2 (-)}^{a b c d} (k)
 =   \sum_{\Sigma_1}   \mathbf{T}_{\mp 4}^{\Sigma_1}  \, \breve{p}_\pm^a    \breve{p}_\pm^b    k^c    k^d
 +    \sum_{\Sigma_2}   \mathbf{T}_{\mp 4}^{\Sigma_2}  \, \breve{p}_\pm^a     \breve{p}_\pm^b     \breve{p}_\pm^c     \breve{p}_\mp^d     \, ,  \qquad
\end{eqnarray}
where  $\Sigma_1$  and  $\Sigma_2$   means sum over  all possible permutations of basic vectors.  The coefficients  have a form
\begin{eqnarray}\label{eq:}
\mathbf{T}_{\mp 4}^{\Sigma_1}   =   \frac{1}{2^4}   (\breve{p}_\mp)_e   \,   (\breve{p}_\mp)_f  \,  q_g    q_h   \,   T^{e f g h }       \, . \qquad
\mathbf{T}_{\mp 4}^{\Sigma_2}   =   \frac{1}{2^4}   (\breve{p}_\mp)_e   \,   (\breve{p}_\mp)_f  \,   (\breve{p}_\mp)_g     (\breve{p}_\pm)_h   \,   T^{e f g h }       \,  . \qquad
\end{eqnarray}
If we count numbers of  basic vectors $\breve{p}_\pm^a$  we can conclude that   relation (\ref{eq:Hln})  $n_\pm +  n_\mp = \lambda$    valid  since $ \pm 2 + 0 = \pm 2$  and $\pm 3 + (\mp 1) = \pm 2$.

We are  interested  in  tensor antisymmetric in the first pair  ($a b$)  as well as  in the second  pair of indices ($c d$).  Then   only  terms of the form
\begin{eqnarray}\label{eq:}
 \mathbf{T}_{\mp 4}^{\Sigma_1}   \to   \mathbf{T}_{\mp 4}^R
 =   \frac{1}{2^4}   \Big(  q_e    (\breve{p}_\mp)_f  - q_f  (\breve{p}_\mp)_e       \Big) \,  \Big(  q_g     (\breve{p}_\mp)_h -  q_h    (\breve{p}_\mp)_g    \Big) \,
 \,   T^{e f g h}      \, , \qquad
\end{eqnarray}
 survive  and we obtain
\begin{eqnarray}\label{eq:Tnmj45}
T_{\pm 2 (-)}^{[a b] [c d] } (k)
 =       \mathbf{T}_{\mp 4}^R  \, \Big(  k^a  \breve{p}_\pm^b    -   k^b  \breve{p}_\pm^a    \Big)
  \Big(  k^c  \breve{p}_\pm^d   -   k^d  \breve{p}_\pm^c    \Big)      \, .  \qquad
\end{eqnarray}
This tensor is  symmetric  under  exchange of the two pairs of indices  ($a b$)  and   ($c d$).
We can rewrite it  in  terms of second rank symmetric tensor  $ h_\pm^{a b}    =   \mathbf{T}_{\mp 4}^R  \breve{p}_\pm^a   \breve{p}_\pm^b $  as
\begin{eqnarray}\label{eq:Tnmj45p}
T_{\pm 2 (-)}^{[a  b] [c d ] } (k)
 =      k^{a}   k^{c} h_\pm^{b d}  -   k^{a}   k^d   h_\pm^{b c}    - k^b   k^{c}  h_\pm^{a d}  +   k^b   k^d   h_\pm^{a c}     \,   .    \qquad
\end{eqnarray}
It is easy to check   symmetry properties of    tensor   $T_{\pm 2 (-)}^{[a b ] [c d ] } (k)$:  cyclic permutation symmetry  and   Bianchi identity
\begin{eqnarray}\label{eq:cpBi}
 T_{\pm 2 (-)}^{a [b c d]}   =  0  \,  ,   \qquad   T_{\pm 2 (-)}^{a b [c d} k^{e ]}  =  0  \, .
\end{eqnarray}

With the help of expressions  for scalar product of the basic vectors  and    (\ref{eq:Tnmj45})   we obtain
\begin{eqnarray}\label{eq:TRicc}
T_{\pm 2 (-)}^{a c}  (k) =   \eta_{b d} T_{\pm 2 (-)}^{a b c d} (k) =    0   \, ,  \qquad   T_{\pm 2 (-)} (k) =  \eta_{a b}  T_{\pm 2 (-)}^{a b}  (k) =  0 \, .
\end{eqnarray}
Using the gauge transformations of the basic vectors  $  \delta  k^a =  0$    and  $ \delta   \breve{p}_\pm^a  =   \varepsilon_\pm   k^a  $  we  have
\begin{eqnarray}\label{eq:GiT2}
\delta  T_{\pm 2 (-)}^{[a b] [c d]  }  (k) =  0  \, ,
\end{eqnarray}
which means that  fields  $T_{\pm 2 (-)}^{[a b ] [c d ] } (k)$   are   gauge invariant   and we can use them  as a physical fields.
They  are irreducible representations of  proper Poincare  group.

\subsubsection{Curvatures  as irreducible representations of  Poincare  group with parity  }

As explained earlier,  instead of two components $  T_{+ 2 (-)}^{[a b ] [c d]}  (k )$  and $  T_{- 2 (-)}^{[a b ] [c d]}  (k )$  we will introduce  tensor  $R^{a b c d} (k) $ which  is irreducible representation of Poincare group with  space inversion.    With the help of   (\ref{eq:GSbh}) we have
\begin{eqnarray}\label{eq:GSbh14}
 R^{a b c d} (k)   =   \alpha   T_{+ 2 (-)}^{[a b ] [c d]}  (k )  +  \beta    T_{+ 2 (-)}^{[a b ] [c d]}  (k )
=  k^{a}   k^{c} h^{b d} (k) -   k^{a}   k^d   h^{b c} (k)   - k^b   k^{c}  h^{a d} (k) +   k^b   k^d   h^{a c} (k)   \, ,  \qquad  \nonumber \\
 h^{a b} (k)  =  \alpha  h_+^{a b} (k) +  \beta h_-^{a b} (k)
 =    \alpha  \mathbf{T}_{- 4}^R (k) \breve{p}_+^a   \breve{p}_+^b    +  \beta    \mathbf{T}_{+ 4}^R  (k) \breve{p}_-^a   \breve{p}_-^b            \, .  \qquad   \qquad    \qquad
\end{eqnarray}

The coordinate  representation  for real fields is
\begin{eqnarray}\label{eq:RItR}
R^{a b c d  } (x) = \int d^4 p  \Big(  R^{a b c d}  (p) +   R^{a b c d} ( - p)  \Big)  e^{- i p x}     \, ,  \qquad
\end{eqnarray}
and with the help of   (\ref{eq:Om2cr})
\begin{eqnarray}\label{eq:RIt}
R^{a b c d  } (x)
=  \partial^{a}   \partial^{c} h^{b d}  (x)  -  \partial^{a}   \partial^d   h^{b c} (x)  -   \partial^b   \partial^{c}  h^{a d}  (x)  + \partial^b   \partial^d   h^{a c}  (x)   \, .  \qquad
\end{eqnarray}
where according to   (\ref{eq:om1x})
\begin{eqnarray}\label{eq:om2xR}
h^{a b }  (x)  =   -  \int  d^4 p  \Big[ h^{a b }   (p) + h^{a b }   (- p)  \Big]  e^{- i p x}  \, .  \qquad
\end{eqnarray}
In the  expression   (\ref{eq:RIt})  we can recognize  Riemann tensor see Ref.\cite{LL},   and in the expression  (\ref{eq:om2xR}) metric tensor in  weak field approximation.

Then we can calculate  linearized  Ricci tensor
\begin{eqnarray}\label{eq:Ricct}
R^{b d} (x)  =  \eta_{a c}  R^{a b c d  } (x)
=  \partial^2  h^{b d}  (x) -  \partial_a   \partial^d   h^{b a}  (x) -     \partial^b   \partial_a   h^{a d}  (x) + \partial^b   \partial^d   h (x)    \, ,  \qquad
\end{eqnarray}
linearized scalar curvature
\begin{eqnarray}\label{eq:}
R  (x)  = \eta_{b d} R^{b d}   =  \partial^2  h  -  \partial_a   \partial_b   h^{b a}   -    \partial_a   \partial_b   h^{a b}  +   \partial^2   h
= 2  \Big(  \partial^2  h  - \partial_a   \partial_b   h^{b a}    \Big)     \, ,  \qquad
\end{eqnarray}
and  linearized Einstein tensor
\begin{eqnarray}\label{eq:EInt}
G^{b d} (x)  = R^{b d} -  \frac{1}{2}  \eta^{b d} R
= \partial^2  h^{b d}  -  \partial_a   \partial^d   h^{b a}   -     \partial^b   \partial_a   h^{a d} +  \partial^b   \partial^d   h
- \eta^{b d}   \Big(  \partial^2  h  - \partial_a   \partial_b   h^{b a}    \Big)  \, .  \qquad
\end{eqnarray}

\subsubsection{Einstein equations in vacuum   }

Using expressions   (\ref{eq:cpBi}),   (\ref{eq:GiT2})    and   (\ref{eq:GSbh14}),  and substituting $k^a \to p^a \to i \partial^a$   we can conclude  that  tensor  $R^{a b c d} (x) $ has standard symmetry properties of Riemann  tensor:
cyclic permutation symmetry  and   Bianchi identity
\begin{eqnarray}\label{eq:}
 R^{a [b c d]}  (x) =  0  \,  ,   \qquad   R^{a b [c d} k^{e ]}  (x) =  0  \, ,
\end{eqnarray}
and that it  is gauge invariant
\begin{eqnarray}\label{eq:}
\delta R^{a b c d}  (x)  = 0  \, . \qquad
\end{eqnarray}
So, it can be used as physical field.

From    (\ref{eq:TRicc})   we obtain  that Ricci and scalar curvatures  vanish
\begin{eqnarray}\label{eq:}
R^{a c}  (x) =   \eta_{b d} R^{a b c d} (x) =    0   \, ,  \qquad   R (x) =  \eta_{a b}  R^{a b}  (x) =  0 \, ,
\end{eqnarray}
which produces  Einstein equations in vacuum
\begin{eqnarray}\label{eq:Eev1}
G^{a c}  (x) =   R^{a c}  (x)  - \frac{1}{2}   \eta^{a c}  R (x)  =  0  \, .
\end{eqnarray}

To express  coupling of  gravitational  fields with   matter (to a conserved energy-momentum tensor) we must solve  Bianchi identity in terms of symmetric tensor $h^{a b} (x)$.
The solution is Eq. (\ref{eq:RIt})  and then Einstein tensor  takes the form (\ref{eq:EInt}).  Consequently,    interaction with  matter can be express as in  (\ref{eq:Inmhab}).

\subsubsection{Dimensions of the  representations and  degrees of freedom   }

Let us count  dimensions of these  representations.   As we have already shown dimension of each of representation $T_{\pm 2 }^{a b c d } (k) $   defined in  Eq.(\ref{eq:NHhg})    is $d_2 = 28$,
where we used the fact  that subspace corresponding to  $\pi_0$  is two dimensional.  The first term   is $6 \cdot 2 \cdot 2 = 24$  dimensional  where $6$ is number of combination of $4$ things taken $2$ at a time and $2 \cdot 2$ is  dimension of  subspace corresponding to two operators $\pi_0$.  The second term is   $4$ dimensional so that $24 + 4 = 28$. After   substitution two dimensional
projector $\pi_0 $ with one dimensional   $\pi_{0 -}$  we obtain that  $T_{\pm 2 (-)}^{a b c d} (k)$, defined in  Eq.(\ref{eq:NHh})  is  $6 \cdot 1 \cdot 1 + 4 = 10$ dimensional representation, which could be expected for symmetric second ranktensor.

Using symmetry properties the second term in (\ref{eq:NHh}) disappears and we are left with $6$ dimensional representation, describing with  Eq.(\ref{eq:Tnmj45}). Since in that equation  we are missing terms  with
$ k^a   k^b $  and $ k^c   k^d $,   we are left with $4$ dimensional representation.  In Eq.(\ref{eq:Tnmj45}) both multipliers are antisymmetric. The other three possibilities
(that both multipliers are symmetric and that one is symmetric and the other is antisymmetric) complements dimension  of such kind of representations  to $4$.
It means that in our particular case   each of  representation  $T_{+ 2 (-)}^{[a b] [c d] } (k)$  and $T_{- 2 (-)}^{[a b] [c d] } (k)$    is one dimensional. Consequently,
they are irreducible representations of  proper Poincare group with one degree of freedom.

According to discussion above, Riemann tensor  $R^{a b c d} (k) $ which  is irreducible representation of Poincare group with  space inversion,  has two  degrees of freedom.

%%%%%%

\section{Massless  tensors with arbitrary rank   }\label{sec:Tar}
\setcounter{equation}{0}

We are going to solve  most general  equation for integer helicity
\begin{eqnarray}\label{eq:}
(S^A{}_B - \lambda \sigma \delta^A_B  )  \Psi^B  = 0  \, ,
\end{eqnarray}
where $\Psi^A$  is tensor with  rank   $n$ so that  $\Psi^A = T^{a_1 a_2 \cdots a_n}$.
The problem comes down to that of  vector fields.

\subsection{Spin operator,  delta function  and  projectors    for  arbitrary rank tensors in terms of elementary projectors   }

First  we are going to express spin operator   $S^A{}_B$  and  delta function  $\delta^A_B$  for fields with  arbitrary helicity   in terms of corresponding  expressions for vector fields.

We will use abbreviated notation (we will omit the indices ).  It means that formally  we multiply first term with first term and  second term with second term
\begin{eqnarray}\label{eq:}
 (A \ast B )  (C \ast D ) =  (A C \ast B D) \, .
\end{eqnarray}

In abbreviated notation  we can write
\begin{eqnarray}\label{eq:S12ara}
 (S_{1 2})^A{}_B
=   \Big( \rho   \delta^{n - 1} +  \delta      \rho \delta^{n- 2}   +  \cdots    \delta^{n - 1}  \rho   \Big)^A{}_B
= \sum_{k = 0}^{n - 1}  \Big(   \delta^k      \rho \delta^{n- k -1}   \Big)^A{}_B     \,   .     \qquad
\end{eqnarray}
Using  relations  $\rho = \pi_+ - \pi_- $   and  $\delta =  \pi_0 + \pi_+ + \pi_- $   we   obtain
\begin{eqnarray}\label{eq:cSrd}
 (S_{1 2})^A{}_B
= \sum_{k = 0}^{n - 1}  \Big(  ( \pi_0 + \pi_+ + \pi_-    )^k  \,  ( \pi_+ - \pi_-   )  \,   ( \pi_0 + \pi_+ + \pi_-  )^{n- k -1}   \Big)^A{}_B     \,   ,     \qquad
\end{eqnarray}
and   after multiplications
\begin{eqnarray}\label{eq:cSrdfa}
 (S_{1 2})^A{}_B   =  \sum_{k = 1}^n   k \Big(   P_k - P_{- k}    \Big)^A{}_B    \,   .     \qquad
\end{eqnarray}

Delta function for tensors with arbitrary rank  has the form
\begin{eqnarray}\label{eq:Defga}
\delta^A_B  =  \delta^n  =   ( \pi_0 + \pi_+ + \pi_- )^n   =  \sum_{k = 1}^n    \Big(  P_k + P_{- k}    \Big)   +  P_0             \, .
\end{eqnarray}

Here $(P_k)^A{}_B$  is defined as the sum of all possible terms $\pi_{i_1}   \pi_{i_2} \cdots   \pi_{i_n}  $ such that $\sum_{m=1}^n i_m = k$   where   we   assign   $\pi_+  \equiv \pi_1$  and 
 $\pi_-  \equiv  \pi_{- 1}$.  Note that  the sequence of multipliers is significant.
Since  $\pi_0$,  $\pi_+$   and  $\pi_-$ are projectors it follows that  all expressions   $P_k$   are  also  projectors,  satisfying the orthogonality condition
  $P_k P_q   =  \delta_{k q } P_k$   where $  (k, q = 0,  \pm 1, \pm 2, \cdots,   \pm n) $.
The multiplication factor  $k$  in (\ref{eq:cSrdfa})  is consequence of combinatorics.

We already had  explicit expressions for   projection  operators in the simplest cases:  for vectors
$ ( P_{ 0} )^a{}_b  =     (\pi_{ 0} )^a{}_b   \, ,  ( P_{\pm 1} )^a{}_b  =     (\pi_{ \pm} )^a{}_b $  and  for second rank tensors  in Eq. (\ref{eq:PO02}), which are particular cases of
projectors    for  arbitrary rank tensors.

\subsection{Solution of spin equation  for arbitrary rank  }

To obtain spectrum of  helicities  we are going to solve consistency condition
\begin{eqnarray}\label{eq:Smlad}
\det ( S_{1 2}   - \lambda \sigma )^A{}_B  =  0      \,   .    \qquad
\end{eqnarray}
Using   (\ref{eq:cSrdfa})  and   (\ref{eq:Defga})    we obtain
\begin{eqnarray}\label{eq:Smlaara}
 (S_{1 2})^A{}_B   - \lambda \sigma \delta^A_B
=    \sum_{k = - n}^n    (k - \lambda \sigma)  ( P_k)^A{}_B             \,   .    \qquad
\end{eqnarray}
Since   $( P_k )^A{}_B$  are  projection  operators  $S_{1 2} - \lambda \sigma$  is diagonalized  and we have
\begin{eqnarray}\label{eq:Smladen}
\det ( S_{1 2} - \lambda \sigma)
=     \prod_{k = - n}^n   ( k - \lambda \sigma)^{d_k}   \det  P_k     = 0         \,   ,    \qquad
\end{eqnarray}
where $d_k$ is dimension of subspace after projection  with $P_k$.

With the help of relations  $\sigma^2 = 1$ and    $d_k = d_{- k}$   we obtain
\begin{eqnarray}\label{eq:Smlade12n}
\det  ( S_{1 2} - \lambda \sigma  \delta)
=   ( - \lambda \sigma)^{d_0}   \prod_{k =1}^n   ( k^2 - \lambda^2 )^{d_k}   \,    \prod_{k = - n}^n   \det  P_k   = 0         \,   .    \qquad
\end{eqnarray}

Using the fact that $\det P_k \ne 0$   and  $\sigma \ne 0$  we can conclude   that
\begin{eqnarray}\label{eq:}
( - \lambda   )^{d_0}   \prod_{k =1}^n   ( k^2 - \lambda^2 )^{d_k}       = 0      \,   ,   \qquad
\end{eqnarray}
which produces eigenvalues of operator   $( S_{1 2})^A{}_B$    and consequently all possible helicities  for $n$ rank tensor
\begin{eqnarray}\label{eq:}
\lambda_0 = 0 \, , \qquad   \lambda_k = k \, , \qquad  \lambda_{- k} = -k     \, ,   \qquad  k = (1, 2, \cdots, n )    \,  .
\end{eqnarray}

The  eigenfunctions, which are our candidates for  irreducible representations,  for massless  tensors  with arbitrary rank  are
\begin{eqnarray}\label{eq:Eifn}
\Psi_m^A   =  (P_m)^A{}_B    \Psi^B    \, .  \qquad    (m = 0, \pm 1, \pm 2 , \cdots , \pm n  )
\end{eqnarray}

\subsection{Dimension of subspaces $d_k$   }

The subspace projected by   $\pi_0$  has dimension $2$ while those projected by  
$\pi_+$  and $\pi_-$  each have dimension $1$. 
Since  $d_k = d_{- k}$  we obtain  following expressions for the dimensions  $d_k$.

If  $P_k$ contain    $m$   operators $\pi_0$ then   for  even $m$  and  $r = \frac{1}{2} (n + k) $,    $s = \frac{m}{2}    $
\begin{eqnarray}\label{eq:}
  d_k  =    \sum_{s = 0}^{r -k}   \frac{n!}{ (r - s)! (r -k - s   )! ( 2 s)!  }  2^{2 s}       \, ,
\end{eqnarray}
and  for   odd  $m$  and  $r = \frac{1}{2} (n + k - 1) $,   $s = \frac{m - 1}{2}    $
\begin{eqnarray}\label{eq:}
 d_k   =   \sum_{s = 0}^{r -k}   \frac{n!}{ (r - s)! (r -k - s   )! ( 2 s + 1)!  }  2^{2 s + 1}       \, .
\end{eqnarray}

In particular we have
\begin{eqnarray}\label{eq:}
 d_n  =    1        \, ,   \qquad
 d_{n - 1}    = 2 n        \, , \qquad      \nonumber \\
d_{n - 2}  =   n (2 n - 1)  \, , \qquad
 d_{n - 3}   =   n (n - 1)   \frac{4 n - 2}{3}     \, .   \qquad
\end{eqnarray}
In next two sections we will consider the   highest   helicity which is always one dimensional.

%%%%%%

\section{ Components with highest  helicity   }
\setcounter{equation}{0}

In previous sections, we demonstrated that our principle  field equations successfully reproduce all known cases with lower helicities   $\lambda = 0, 1, 2  $.  
We now turn to a comparison of our framework with established results for higher helicity fields  
 $\lambda \ge 3$. 

The general theory of free fields with arbitrary helicity $n$ was first developed by Fronsdal \cite{Fr}. We will show that Fronsdal's formulation emerges as a special case of our approach when restricted to the highest helicity component.

\subsection{Gauge transformation for massless  tensors of arbitrary rank  }

To find  little group transformation for massless  tensors of arbitrary rank we will use little group transformation for massless  vector  fields.
With the help of   (\ref{eq:BEder})  in   abbreviated notation  we have
\begin{eqnarray}\label{eq:}
  (\Pi_1 )^A{}_B  =   \sum_{k = 0}^{n - 1}  \Big(   \delta^k     \hat{\Pi}_1 \delta^{n- k -1}   \Big)^A{}_B        \, ,  \qquad
   (\Pi_2 )^A{}_B  =   \sum_{k = 0}^{n - 1}  \Big(   \delta^k     \hat{\Pi}_2 \delta^{n- k -1}   \Big)^A{}_B              \, ,
\end{eqnarray}
where $(\hat{\Pi}_1)^a{}_b$  and  $(\hat{\Pi}_2)^a{}_b$ in the sum are expressions for vector fields and $\delta^k$ is product of $k$  corresponding  $\delta$ functions.

Tensor fields with helicity  $\lambda_i$  has a form
\begin{eqnarray}\label{eq:GThl}
 T_{\pm i}^{a_1  a_2 \cdots a_n}   ( k)
=    (P_{\pm i})^{a_1  a_2 \cdots a_n}{}_{b_1  b_2 \cdots b_n}    T^{b_1  b_2 \cdots b_n}  (k)    \,  .   \qquad
  (i = 0,  1,  2 , \cdots ,  n  )
\end{eqnarray}
 With help of (\ref{eq:Txyvg}) we can obtain its  gauge transformations
\begin{eqnarray}\label{eq:dTpmi}
\delta  T_{\pm i}^{a_1  a_2 \cdots a_n}   ( \varepsilon_1,  \varepsilon_2, k)
= \sum_{k = 0}^{n - 1} \Big( \delta^k  \hat{\Delta}  \delta^{n - k -1}     \Big)^{a_1  a_2 \cdots a_n}{}_{b_1  b_2 \cdots b_n}    T_{\pm i}^{b_1  b_2 \cdots b_n}  (k)   \,  ,   \qquad
\end{eqnarray}
where $ \hat{\Delta}^a{}_b   =  i \Big(  \varepsilon_1 \hat{\Pi}_1  +  \varepsilon_2  \hat{\Pi}_2    \Big)^a{}_b  $  is  expression for vector fields.

Using expressions for eigenfunctions  (\ref{eq:GThl})  and for projection operators as well as  for  transformation of basic  vector fields (\ref{eq:gtbv})
and the fact that $\hat{\Delta}^a{}_b$  acts on vectors and not on scalars we  can obtain gauge transformation of  arbitrary tensor.

The general expression   for highest  helicity  tensor  (for example tensor of rank  $n$ with helicities $\lambda_{\pm n}  = \pm n$)  has a form
\begin{eqnarray}\label{eq:Hhn}
T_{\pm n}^{a_1  a_2 \cdots a_n}  =    (P_{\pm n})^{a_1  a_2 \cdots a_n}{}_{b_1  b_2 \cdots b_n}    T^{b_1  b_2 \cdots b_n}
=       \mathbf{T}_{\mp n}  \, \breve{p}_\pm^{a_1}    \breve{p}_\pm^{a_2}  \cdots \breve{p}_\pm^{a_n}     \, , \qquad
\end{eqnarray}
where
\begin{eqnarray}\label{eq:}
 \mathbf{T}_{\mp n}  =  \frac{(-1)^n}{2^n}  (\breve{p}_\mp)_{b_1}  \,   (\breve{p}_\mp)_{b_2} \, \cdots \,  (\breve{p}_\mp)_{b_n}  \,   T^{b_1  b_2 \cdots b_n}   \, . \qquad
\end{eqnarray}

If  $T^{a_1  a_2 \cdots a_n}$ is    antisymmetric in any two indices then  $ \mathbf{T}_{\mp n} $  and consequently    $T_{\pm n}^{a_1  a_2 \cdots a_n}$    go to zero.
Therefore  representation with  highest  helicity is  totally symmetric.

We can understand this fact with the help of basic vectors.
The  basic vectors $ \breve{p}_\pm^a$   carry  helicities  $\pm 1$ and both basic vectors $k^a$  and $q^a$   carry  helicities  $0$.
If $n_\pm$  are numbers of vectors  $ \breve{p}_\pm^a$  in the representation of field  $\Psi^A$,   then its helicity is
\begin{eqnarray}\label{eq:Hlnl}
\lambda = n_+ - n_-  \, .
\end{eqnarray}
So,  representation with  highest  helicity contains all basic vectors  $ \breve{p}_+^a$   or all basic vectors  $ \breve{p}_-^a$  and obviously is totally symmetric.

Its gauge transformation is
\begin{eqnarray}\label{eq:Hhn1}
\delta T_{\pm n}^{a_1  a_2 \cdots a_n}  =     \sum_{i = 1}^n   k^{a_i}    \Omega_{\pm n}^{a_1  a_2 \cdots  a_{i - 1} a_{i+1}a  \cdots _n} \, , \qquad
\end{eqnarray}
where  gauge parameter has the form
\begin{eqnarray}\label{eq:Hhn1p}
 \Omega_{\pm n}^{a_1  a_2 \cdots  a_{i - 1} a_{i+1}a  \cdots _n}
=      \mathbf{T}_{\mp n}  \, \varepsilon_\pm   \breve{p}_\pm^{a_1}    \breve{p}_\pm^{a_2}  \cdots \breve{p}_\pm^{a_{i - 1}}  \breve{p}_\pm^{a_{i + 1}}  \cdots   \breve{p}_\pm^{a_n}   \, , \qquad
\end{eqnarray}

\subsection{ Highest  helicity  tensor $h_A (k)$ symmetric under space inversion  in the frame of  standard momentum  }

From now on we will use the simplified notation.  With capital $A$ we will denote set of all indices $A = \{a_1  a_2 \cdots a_n \} $. The set of all indices except $a_i$ we will  denote  as
$A_i = \{a_1  a_2 \cdots a_{i - 1}  a_{i + 1}  \cdots    a_n \} $  and similarly  set of all indices except $a_i$  and  $a_j$ we will  denote  as $A_{i j}$.

If the tensor  is symmetric under space inversion then  instead  of two components $T^A_{+ n}  (k)$ and   $T^A_{- n}  (k)$   which are incomplete irreducible representation
of proper Poincare group we will introduce one vector field   $h^A (k)  $  which is incomplete irreducible representation of wider group which includes space inversion. Then  we obtain
\begin{eqnarray}\label{eq:}
h^A   (k) =   \alpha  T_{+ n}^A   (k)  +  \beta T_{- n}^A    (k)  \,  ,
\end{eqnarray}

We also introduce one parameter  $\Omega^{A_i }  (k) =   \alpha  \Omega_{+ n}^{ A_i }  (k)  +  \beta \Omega_{- n}^{ A_i }   (k)  $
instead of two components  $\Omega_{+ n}^{A_i }   (k)$   and  $\Omega_{- n}^{A_i }   (k)  $  related by space inversion.

The field  $h^A (k)  $     is symmetric in all indices
\begin{eqnarray}\label{eq:hConn}
 h_{a_a a_2 \cdots    a_n}  =    h_{( a_a a_2 \cdots    a_n )}    \, ,
\end{eqnarray}
and  using  $\eta^{a_1 a_2}  \breve{p}_\pm^{a_1}    \breve{p}_\pm^{a_2} = 0 $   and  $k_a    \breve{p}_\pm^a = 0 $   we conclude that
\begin{eqnarray}\label{eq:hCon}
 \eta^{a b} h_{a b  A_{i j} }  (k)   =  0    \, ,    \qquad   k^a  h_{a A_i}    (k) =  0   \, .
\end{eqnarray}
The  gauge  transformation  of  field  $ h_A   (k)$   has the  form
\begin{eqnarray}\label{eq:GThA}
\delta h_A  (k) =    \sum_{i = 1}^n  k_{a_i}  \Omega_{A_i}   (k)  \, . \qquad
\end{eqnarray}
With the help of   expression  $\eta^{a_1 a_2}  \breve{p}_\pm^{a_1}    \breve{p}_\pm^{a_2} = 0 $ and using   (\ref{eq:Hhn1p})  we also have condition on parameter $\Omega$
\begin{eqnarray}\label{eq:Frcp}
 \eta^{a b}  \Omega_{ a b  A_{i j k} }  (k) =  0     \, . \qquad  (n \ge  3)
\end{eqnarray}

%%%%%%%

\section{Relations with previous work  }
\setcounter{equation}{0}

In this section, we employ our framework to systematically recover fundamental results from established studies of highest-helicity fields.

\subsection{ The highest-helicity  fields   in coordinate representation  necessarily satisfy the Fierz-Pauli constraints  }

Let us   go from standard momentum frame  to arbitrary frame  and then to coordinate  dependent   fields,  $k^a  \to p^a \to    i \partial^a$.
Then we will consider field  $h^A (x) $
\begin{eqnarray}\label{eq:HaTpm}
h^A (x) =  \int d^4 p  \Big( h^A (p)  +  h^A  (- p)   \Big) e^{- i p x }   \nonumber \\
=  \int d^4 p  \Big[   \alpha  \Big(   T^A_{+ n}  (p )   +   T^A_{+ n}    (- p )   \Big)  +   \beta \Big(   T^A_{- n}   (p )   +   T^A_{- n}   (- p )      \Big)     \Big] e^{- i p x }     \, ,
\end{eqnarray}
and  gauge  parameters  $ \Omega^{A_i} (x)$   in coordinate representation
\begin{eqnarray}\label{eq:HaTpmn}
\Omega^{A_i} (x) =  \int d^4 p  \Big( \Omega^{A_i}  (p)  +  \Omega_{A_i}  (- p)   \Big) e^{- i p x }   \nonumber \\
=  \int d^4 p  \Big[   \alpha  \Big(   \Omega^{A_i}_{+ n}  (p )   +   \Omega^{A_i}_{+ n}    (- p )   \Big)  +   \beta \Big(  \Omega^{A_i}_{- n}   (p )   +   \Omega^{A_i}_{- n}   (- p )      \Big)     \Big] e^{- i p x }     \, .
\end{eqnarray}

The fields   satisfy  massless Klein-Gordon equation and has  following   gauge  transformation
\begin{eqnarray}\label{eq:}
\partial^2  h_A   (x) = 0   \, ,  \qquad
\delta h_A  (x) =    \sum_{i = 1}^n  \partial_{a_i}  \Omega_{A_i}   (x)  \, , \qquad
\end{eqnarray}
when gauge  parameters  according to    (\ref{eq:Frcp})   satisfy    Fronsdal condition \cite{Fr}
\begin{eqnarray}\label{eq:FcOm}
 \eta^{a b}  \Omega_{ a b  A_{i j k}}  (x) =  0     \, . \qquad  (n \ge  3)   \, .
\end{eqnarray}
Later we will see that this is necessary condition that equation of motion is gauge invariant.

As  well  as fields  $T^A_{\pm n} $ the field $h_A$   is totally symmetric  and  satisfy the same  supplementary conditions
\begin{eqnarray}\label{eq:}
  h_{a_a a_2 \cdots    a_n} (x) =    h_{( a_a a_2 \cdots    a_n )}  (x)  \, ,    \qquad     \eta^{a b} h_{a b A_{i j}}  (x)   =  0    \, ,    \qquad   \partial^a  h_{a A_i}    (x) =  0   \, .   \qquad
\end{eqnarray}
These are exactly    Fierz-Pauli  constraints    \cite{FP},  which   eliminate the unwanted   components  ($\lambda < n$) of the tensor   field  $h_A (x)$.

\subsection{ Fronsdal approach  }

In this subsection we will derive Fronsdal equations and  Fronsdal  action  \cite{Fr}.  The corresponding  field consist of two   components with highest helicity, which are one dimensional.
Consequently the field  has two degrees of freedom.
To satisfy that requirement we must impose a condition on  the field to be double-traceless.

\subsubsection{ Derivation of  Fronsdal equations  }

Consequently,     we obtain   equations for tensor field $h_A$ with helicity $\lambda = n$  in arbitrary  frame
\begin{eqnarray}\label{eq:con11c1}
\partial^2  h_A   (x) = 0   \, ,  \qquad     \partial^a  h_{a A_i}    (x) =  0  \, ,    \qquad  \eta^{a b} h_{a b A_{i j}}  (x)     =  0   \, .  \qquad
\end{eqnarray}
This shows  that field  equations  consists of  massless Klein-Gordon  equation  plus supplementary conditions (gauge fixing).
Supplementary conditions define  the minimal set of auxiliary fields suitable for the description of the action    found in  Refs.\cite{SH}  and \cite{Fr}.

Let us combine two conditions from  (\ref{eq:con11c1})  into one condition known as   de Donder gauge
\begin{eqnarray}\label{eq:}
 G_{A_i}  =    \partial^b   h_{b A_i}  - \frac{1}{2}   \sum_{j=1 \, j \ne i}^n    \partial_{a_j}     h^b{}_{b  A_{i j} }  = 0    \, .    \qquad
\end{eqnarray}

Using expression    (\ref{eq:GThA})    we can find particular  gauge transformations
\begin{eqnarray}\label{eq:}
\delta h_A  (k) =    \sum_{i = 1}^n  \partial_{a_i}  \Omega_{A_i}   (k)  \, , \qquad
\delta h_{b A_i}  =    \partial_b   \Omega_{A_i}    +    \sum_{j = 1 \, j \ne i}^n  \partial_{a_j}  \Omega_{b A_{i j}}     \, , \qquad     \nonumber \\
\delta h^b{}_{b A_{i j} }  =    2  \partial^b   \Omega_{b A_{i j} }   +   \sum_{k = 1 \,  k\ne i, j}^n  \partial_{a_k}  \Omega^b{}_{b A_{i j k}} +   \sum_{k = 1 \,  k\ne i, j}^n  \partial_{a_k}  \Omega^b{}_{b A_{i j k}}
 =    2  \partial^b   \Omega_{b A_{i j} }  +   \sum_{k = 1 \,  k\ne i, j}^n  \partial_{a_k}  \Omega^b{}_{b A_{i j k}}    \, .  \qquad
\end{eqnarray}

Then  gauge variation of the de Donder tensor is
\begin{eqnarray}\label{eq:}
 \delta   G_{A_i}  =    \partial^b   \Big(    \partial_b   \Omega_{A_i}    +    \sum_{j = 1 \, j \ne i}^n  \partial_{a_j}  \Omega_{b A_{i j}}    \Big)
  - \frac{1}{2}   \sum_{j=1 \, j \ne i}^n    \partial_{a_j} \Big(    2  \partial^b   \Omega_{b A_{i j} }  +   \sum_{k = 1 \,  k\ne i, j}^n  \partial_{a_k}  \Omega^b{}_{b A_{i j k}}    \Big)   \nonumber \\
  =   \partial^2    \Omega_{A_i}    - \frac{1}{2}   \sum_{j=1 \, j \ne i}^n   \sum_{k = 1 \,  k\ne i, j}^n    \partial_{a_j}   \partial_{a_k}  \Omega^b{}_{b A_{i j k}}    \, .    \qquad
\end{eqnarray}

 Equations (\ref{eq:con11c1}) can be combined  into  one equation  as linear combination of  massless Klein-Gordon   equation and supplementary conditions in terms of  de Donder gauge
\begin{eqnarray}\label{eq:}
\partial^2 h_A    +  \sum_{i = 1}^n  a_i  \partial_{a_i}  G_{A_i}   = 0   \, .    \qquad
\end{eqnarray}

We can find  coefficients $a_i$,    from requirements  that this  equation is  gauge invariant  which means that
\begin{eqnarray}\label{eq:}
\partial^2   \sum_{i = 1}^n  \partial_{a_i}  \Omega_{A_i}
+  \sum_{i = 1}^n  a_i  \partial_{a_i} \Big(   \partial^2    \Omega_{A_i}   - \frac{1}{2}   \sum_{j=1 \, j \ne i}^n   \sum_{k = 1 \,  k\ne i, j}^n    \partial_{a_j}   \partial_{a_k}  \Omega^b{}_{b A_{i j k}}   \Big)  = 0   \, .    \qquad
\end{eqnarray}
This is satisfied for    $a_i = - 1$   and
\begin{eqnarray}\label{eq:}
 \sum_{i = 1}^n \sum_{j=1 \, j \ne i}^n   \sum_{k = 1 \,  k\ne i, j}^n   \partial_{a_i}  \partial_{a_j}   \partial_{a_k}  \Omega^b{}_{b A_{i j k}}  = 0 \, .
\end{eqnarray}

Consequently,  in order to ensure that the Fronsdal equation is gauge invariant, we must  require  that the gauge parameter is traceless
\begin{eqnarray}\label{eq:}
 \Omega^b{}_{b A_{i j k}}  = 0   \,  .
\end{eqnarray}
In our approach, this condition is  satisfied,  as we have shown in   (\ref{eq:FcOm}).

So, equation of motion takes the form  of  Ricci like tensor
\begin{eqnarray}\label{eq:RIltn}
 R_A (h)  \equiv   \partial^2 h_A    -  \sum_{i = 1}^n   \partial_{a_i}  G_{A_i}
=  \partial^2 h_A    -  \sum_{i = 1}^n    \partial_{a_i}   \partial^b   h_{b A_i}    + \frac{1}{2}   \sum_{i=1}^n    \sum_{j=1 \, j \ne i}^n   \partial_{a_i} \partial_{a_j}      h^b{}_{b  A_{i j} }   = 0   \, .    \qquad
\end{eqnarray}

\subsubsection{ Fronsdal action  }

Then  scalar like tensor  is
\begin{eqnarray}\label{eq:}
R_{A_{i j}} = \eta^{a_i  a_j} R_A
=    \partial^2   h^a {}_{ a A_{i j} }   -  2 \partial^a       G_{ a   A_{i j }}     -  \sum_{i = 1}^n   \partial_{a_i}  G^a{}_{a A_{i j k}}   \nonumber \\
=  2  \partial^2   h^a {}_{ a A_{i j} }   -   2 \partial^a    \partial^b   h_{ a b  A_{i j}}
+    \sum_{k =1 \, k \ne i, j}^n  \partial_{a_k}    \partial^b   h^a{}_{a b  A_{i j k}}   \, ,   \qquad       \qquad
\end{eqnarray}
and   Einstein like tensor
\begin{eqnarray}\label{eq:}
G_A  =   R_A - \frac{1}{2}   \sum_{i=1}^n    \sum_{j=1 \, j \ne i}^n    \eta_{a_i  a_j}  R_{A_{i j}}
=  \partial^2 h_A    -  \sum_{i = 1}^n    \partial_{a_i}   \partial^b   h_{b A_i}    + \frac{1}{2}   \sum_{i=1}^n    \sum_{j=1 \, j \ne i}^n   \partial_{a_i} \partial_{a_j}      h^b{}_{b  A_{i j} }   \nonumber \\
- \frac{1}{2}  \sum_{i=1}^n    \sum_{j=1 \, j \ne i}^n   \eta_{a_i  a_j} \Big(  2  \partial^2   h^a {}_{ a A_{i j} }   -   2 \partial^a    \partial^b   h_{ a b  A_{i j}}
+    \sum_{k =1 \, k \ne i, j}^n  \partial_{a_k}    \partial^b   h^a{}_{a b  A_{i j k}}    \Big)     \, .    \qquad
\end{eqnarray}

So, we can take  the action  in the form
\begin{eqnarray}\label{eq:}
S_n (h)   =  \frac{1}{2}  \int  d^4 x     h^A G_A  =   \frac{1}{2} \int  d^4 x    \Big[  h^A  \partial^2 h_A    -  \sum_{i = 1}^n    h^A  \partial_{a_i}   \partial^b   h_{b A_i}
+ \frac{1}{2}   \sum_{i=1}^n    \sum_{j=1 \, j \ne i}^n   h^A  \partial_{a_i} \partial_{a_j}      h^b{}_{b  A_{i j} }   \nonumber \\
- \frac{1}{2}  \sum_{i=1}^n    \sum_{j=1 \, j \ne i}^n   \eta_{a_i  a_j} \Big(  2  h^A  \partial^2   h^a {}_{ a A_{i j} }   -   2   h^A  \partial^a    \partial^b   h_{ a b  A_{i j}}
+    \sum_{k =1 \, k \ne i, j}^n  h^A    \partial_{a_k}   \partial^b   h^a{}_{a b  A_{i j k}}    \Big)    \Big]   \, .  \qquad \qquad
\end{eqnarray}
Using  the notation   $ h_{ A_{i j} }  \equiv   \eta^{a b}  h_{a b  A_{i j} } =   h^a{}_{a  A_{i j} } $,  up to total derivatives we have
\begin{eqnarray}\label{eq:}
S_n (h)   =   \frac{1}{2} \int  d^4 x    \Big[ - \partial^a  h^A  \partial_a h_A    +  n  \partial_a    h^{a A_i}    \partial^b   h_{b A_i}
+  n (n - 1)   \partial_a   \partial_b   h^{a b A_{i j} }       h^c{}_{c  A_{i j} }   \nonumber \\
+   \frac{n (n - 1)}{2}  \partial^c  h_b{}^{b A_{i j} }  \partial_c   h^a {}_{ a A_{i j} }
+   \frac{n (n - 1)  (n - 2)}{4}      \partial_c    h_b{}^{b c A_{i j k} }     \partial^b   h^a{}_{a b  A_{i j k}}    \Big]   \, .   \qquad
\end{eqnarray}
in complete agreement with Ref.\cite{Fr}.

\subsubsection{Propagating degrees of freedom of Fronsdal field}

From general consideration we  know that highest helicity representation is one dimensional and carry one degree of freedom. Then  field  in Fronsdal equation consists of two highest helicity  components
and   must  carry two degrees of freedom.
 We will assume that it is   $k$-traceless  tensor  and we will find $k$ so that  corresponding tensor    carries two degrees of freedom.  Note that  the condition  $n \ge 2 k$   must be satisfied.

 Completely symmetric  $k$-traceless  tensor  with helicity $n$ has
\begin{eqnarray}\label{eq:hkdA0}
 h^k_A   =  {  n + 3  \choose 3 } -  {  n + 3 - 2 k \choose  3 }    \, , \qquad  \qquad
\end{eqnarray}
independent components.  After sme calculation we obtain
\begin{eqnarray}\label{eq:hkdA}
h^k_A    =  n^2  k   - 2 n k (   k - 2  )  - 4 k^2    +   \frac{k}{3}   \Big( 4 k^2    +  11   \Big)          \, . \qquad  \qquad
\end{eqnarray}

Some of these components are related by the gauge condition  $G_{A_i} = 0 $.  Since  de Donder tensor is traceless,  $\eta^{a b} G_{a b A_{i j k}}  = 0$,     we should imposes \begin{eqnarray}\label{eq:}
\Omega^\prime_{A_i}  =  {  n + 2  \choose 3 } -  {  n  \choose  3 }  =  n^2    \, , \qquad   (n \ge 3) \qquad
\end{eqnarray}
conditions.
Note that  parameter  $\Omega_{A_i}$ has one index less  then field $h_A$.   Consequently comparing with  (\ref{eq:hkdA0})    in the first term we will take  $n \to n - 1$.
Because  parameter  $\Omega_{A_i}$  is  traceless  in the second term we will take   $k = 1$  and   $n \to n - 1 $.

In de Donder gauge, the Fronsdal equation  becomes
\begin{eqnarray}\label{eq:}
\partial^2 h_A       = 0   \, .   \qquad
\end{eqnarray}
Note that de Donder gauge does not fix the gauge completely. Both Fronsdal equation and   de Donder gauge are invariant under  residual gauge transformation  given by   traceless gauge parameters $ \omega_{A_i}$   which obey $ \partial^2    \omega_{A_i} = 0$. This residual gauge kills additional $\omega^\prime_{A_i} = n^2  $  components.

Therefore, since complete number of degrees of freedom  must be equal to two  we have
\begin{eqnarray}\label{eq:}
 h^k_A  -  \Omega^\prime_{ A_i} -  \omega^\prime_{ A_i} =     n^2  k   - 2 n k (   k - 2  )  - 4 k^2    +   \frac{k}{3}   \Big( 4 k^2    +  11   \Big)    -  2 n^2 =  2   \, . \qquad  \qquad
\end{eqnarray}
This is cubic equation for $k$  with  one solution  $k  =  2$.
There is one more solution  with integers $k =3$  and $n = 3$. But it does not satisfy condition $n \ge 2 k$.

This means that double-traceless fields for arbitrary  helicity  in  Fronsdal equation indeed propagates two degrees of freedom.

\subsection{ Relation with  de Wit Freedman  approach   }

Let us shorty comment de Wit and Freedman  approach  \cite{dWF}, based on a hierarchy of generalized Christoffel symbols.  Their final result  is  Fronsdal's action  which we obtained in previous section.
Here we are going to comment  their intermediate result: the gauge invariant generalized curvature.

We are  interested  in  tensors  of the form
\begin{eqnarray}\label{eq:Tnmj45a}
T_{\pm  2 n  (-)}^{(a_1  a_2 \cdots a_n ) |  ( b_1  b_2 \cdots b_n ) } (k)
 =       \mathbf{T}_{\mp 2 n}^R  \, \prod_{i = 1} ^n   \Big(  k^{a_i}  \breve{p}_\pm^{b_i }    -   k^{b_i}  \breve{p}_\pm^{a_i }    \Big)       \, .  \qquad
\end{eqnarray}
Using transformations  $\delta k^a = 0$  and  $\delta   \breve{p}_\pm^a =  \varepsilon_\pm k^a    $     we can conclude  that every term is gauge invariant
$\delta  \Big(  k^{a_i}  \breve{p}_\pm^{b_i }    -   k^{b_i}  \breve{p}_\pm^{a_i }    \Big) = 0 $.   Consequently  tensors  $T_{\pm  2 n  (-)}^{(a_1  a_2 \cdots a_n ) |  ( b_1  b_2 \cdots b_n ) } (k)  $
are  gauge invariant   and we can use them  as a physical fields.   They  are irreducible representations of  proper Poincare  group.

The pair exchange property
\begin{eqnarray}\label{eq:Tnmj45a1}
T_{\pm  2 n  (-)}^{(a_1  a_2 \cdots a_n ) |  ( b_1  b_2 \cdots b_n ) } (k)   = (- 1)^n   T_{\pm  2 n  (-)}^{  ( b_1  b_2 \cdots b_n )  |  (a_1  a_2 \cdots a_n ) } (k)            \, ,  \qquad
\end{eqnarray}
follows from the  fact that  expression $  k^{a_i}  \breve{p}_\pm^{b_i }    -   k^{b_i}  \breve{p}_\pm^{a_i }  $  change the sign after exchange $a_i$ with $b_i$.

Using the fact that
\begin{eqnarray}\label{eq:}
    k^a    \Big(  k^b   \breve{p}_\pm^c     -   k^c  \breve{p}_\pm^b     \Big) +  cyclic  (a, b, c)  =  0       \, ,  \qquad
   \breve{p}_\pm^a    \Big(  k^b   \breve{p}_\pm^c     -   k^c  \breve{p}_\pm^b     \Big) +  cyclic  (a, b, c)  =  0       \, ,  \qquad
\end{eqnarray}
we can  check   symmetry properties:   cyclic permutation symmetry
\begin{eqnarray}\label{eq:Tnmj45a12}
T_{\pm  2 n  (-)}^{(a_1  a_2 \cdots a_n ) |  ( b_1  b_2 \cdots b_n ) } (k)  +  cyclic  (a_i, b_i, b_j)  =  0             \, ,  \qquad
\end{eqnarray}
and   Bianchi identity
\begin{eqnarray}\label{eq:Tnmj45a1b}
k^a T_{\pm  2 n  (-)}^{(a_1  a_2 \cdots a_n ) |  ( b_1  b_2 \cdots b_n ) } (k)  +  cyclic  (a, a_i, b_i )  =  0             \, .  \qquad
\end{eqnarray}

With the help of expressions  for scalar product of the basic vectors  and    (\ref{eq:Tnmj45a})   we obtain
\begin{eqnarray}\label{eq:TRicca}
  \eta_{a_i b_i}   T_{\pm  2 n  (-)}^{(a_1  a_2 \cdots a_n ) |  ( b_1  b_2 \cdots b_n ) } (k)    =    0   \, .
\end{eqnarray}

We can rewrite   (\ref{eq:Tnmj45a})    in terms of $n$ rank symmetric tensor
\begin{eqnarray}\label{eq:}
h_\pm^{ c_1  c_2 \cdots c_n }    (k)
 =       \mathbf{T}_{\mp 2 n}^R  \,   \breve{p}_\pm^{c_1 }     \breve{p}_\pm^{c_2 }   \cdots     \breve{p}_\pm^{c_n }        \, .  \qquad
\end{eqnarray}
as
\begin{eqnarray}\label{eq:Tnmj45ar}
T_{\pm  2 n  (-)}^{(a_1  a_2 \cdots a_n ) |  ( b_1  b_2 \cdots b_n ) } (k)
 =       \sum_{ m= 0}^n   (-1)^{n - m}  k^{a_1}  \cdots k^{a_m}   k^{b_{m +1}}  \cdots   k^{b_n} h_\pm^{ b_1   \cdots   b_m  a_{m +1} \cdots a_n }        \, .  \qquad
\end{eqnarray}

\subsubsection{Generalized  curvatures in coordinate representation  }

As explained earlier,   with the help of  (\ref{eq:GSbh})  instead of two components $ T_{+  2 n  (-)}^{(a_1  a_2 \cdots a_n ) |  ( b_1  b_2 \cdots b_n ) } (k)  $  and   $ T_{-  2 n  (-)}^{(a_1  a_2 \cdots a_n ) |  ( b_1  b_2 \cdots b_n ) } (k)  $    we will introduce   tensor
\begin{eqnarray}\label{eq:GSbh14g}
 R^{a_1  a_2 \cdots a_n,  b_1  b_2 \cdots b_n } (k)    =   \alpha  T_{+  2 n  (-)}^{(a_1  a_2 \cdots a_n ) |  ( b_1  b_2 \cdots b_n ) } (k)
  +  \beta    T_{-  2 n  (-)}^{(a_1  a_2 \cdots a_n ) |  ( b_1  b_2 \cdots b_n ) } (k)       \, ,  \qquad   \qquad    \qquad
\end{eqnarray}
and  instead of two components  $ h_+^{a_1  a_2 \cdots a_n }     (k) $  and  $ h_-^{a_1  a_2 \cdots a_n }    (k)  $  we will introduce   tensor
\begin{eqnarray}\label{eq:GSbh14ga}
h^{a_1  a_2 \cdots a_n }     (k)      =  \alpha   h_+^{a_1  a_2 \cdots a_n }   (k)    +  \beta   h_-^{a_1  a_2 \cdots a_n }     (k)     \, .  \qquad   \qquad    \qquad
\end{eqnarray}
Both  tensors are  irreducible representations of Poincare group with  space inversion.

The coordinate  representations  for real fields are   generalized  Riemann tensor see Ref.\cite{dWF}
\begin{eqnarray}\label{eq:RItRa}
 R^{a_1  a_2 \cdots a_n,  b_1  b_2 \cdots b_n }  (x) = \int d^4 p  \Big(   R^{a_1  a_2 \cdots a_n,  b_1  b_2 \cdots b_n }    (p) +   R^{a_1  a_2 \cdots a_n,  b_1  b_2 \cdots b_n }  ( - p)  \Big)  e^{- i p x}     \, .  \qquad
\end{eqnarray}
and generalized metric tensor
\begin{eqnarray}\label{eq:RItRh}
 h^{a_1  a_2 \cdots a_n }  (x) = \int d^4 p  \Big(   h^{a_1  a_2 \cdots a_n }    (p) +   h^{a_1  a_2 \cdots a_n }  ( - p)  \Big)  e^{- i p x}     \, .  \qquad
\end{eqnarray}

Just like a tensors  $T_{\pm  2 n  (-)}^{(a_1  a_2 \cdots a_n ) |  ( b_1  b_2 \cdots b_n ) } (k)  $  generalized  Riemann tensor $ R^{a_1  a_2 \cdots a_n,  b_1  b_2 \cdots b_n }  (x)  $
 is gauge invariant and   satisfy cyclic permutation symmetry and   Bianchi identity.

For $n = 1$ we obtain Maxwell field strength
\begin{eqnarray}\label{eq:}
R^{a b}  =  \partial^a h^b  -  \partial^b  h^a  \, ,
\end{eqnarray}
and for $n = 2$   Riemann tensor in weak field approximation   (\ref{eq:RIt})
\begin{eqnarray}\label{eq:RIt2}
R^{a b c d  } (x)
=  \partial^{a}   \partial^{c} h^{b d}    -  \partial^{a}   \partial^d   h^{b c}   -   \partial^b   \partial^{c}  h^{a d}    + \partial^b   \partial^d   h^{a c}     \, .  \qquad
\end{eqnarray}

Note that   tensor  $R^{a b c d  }$,  which denoted  in   Ref.\cite{dWF}  with prime,
is symmetric under  exchange  both indices $a b$  with $c d$ and not under separately exchange $a$ with $c$  or $b$  with $d$.
We can construct   symmetric tensor   in both $a c$  and  $b d$  indices   as linear combination of antisymmetric Riemann  tensors in the   same way  as in  equation  (2.10)   of Ref.\cite{dWF}
\begin{eqnarray}\label{eq:RIt2a}
R_S^{a b c d  }  =     \frac{1}{2}     \Big(    R^{a c b d  }   +  R^{a d b  c   }   \Big) \, ,  \qquad
\end{eqnarray}
where $R_S^{a b c d  } $ is   symmetric tensor of Ref.\cite{dWF}  which  in that article  generalizes  to higher spin.

The  generalized Riemann tensor of    Ref.\cite{dWF}  can be obtained after symmetrization  of the tensor  (\ref{eq:RItRa})   with respect of all indices $a_i$  as well as with respect of all indices $b_i$.

%%%%%%%%

\section{ Arbitrary rank tensor with helicity $\lambda = 1$  produces Maxwell equations and with $\lambda = 2$ Einstein equations }
\setcounter{equation}{0}

Arbitrary rank tensor with  $n \ge 2$ components  contains  also   helicities   $\lambda = \pm 1$  and  $\lambda = \pm 2$. We will show   that these  tensors  describe  Maxwell equations  and  Einstein equations in weak field approximation.

\subsection{Arbitrary  rank tensor with  helicities  $\lambda = \pm 1$  }

The components   of  $n$  rank tensor  with  helicities $\lambda  = \pm 1$  have a form
\begin{eqnarray}\label{eq:}
T_{\pm 1}^A   (k)  =  (P_{\pm 1})^A{}_B    T^B (k) =  \sum_{i = 1}^n  (\pi_\pm)^{a_i}{}_{b_i}    (\pi_0 )^{A_i}{}_{B_i}     T^{b_i  B_i}    (k)              \, .
\end{eqnarray}
Since $\pi_0$ project to two dimensional space and   $\pi_+$ and $\pi_-$   to one dimensional space,  both   components $T_{\pm 1}^A   (k)  $  are  $ 2^{n - 1} n $  dimensional representations and they  have
$ 2^{n - 1} n $ degrees of freedom.

To avoid influence of  nontrivial  gauge transformations of $( \pi_{ 0 +} )^a{}_b$   we will save  only $( \pi_{ 0 -} )^a{}_b$ part  (gauge  invariant     part of
$( \pi_{ 0 } )^a{}_b$)   which means  that we will take  $( \pi_{ 0 } )^a{}_b  \to ( \pi_{ 0 -} )^a{}_b$.  With that choice  we obtain
\begin{eqnarray}\label{eq:}
T_{\pm 1 (-)}^A (k) =  (P_{\pm 1 (-)})^A{}_B    T^B   (k)  =   \sum_{i = 1}^n  (\pi_\pm )^{a_i}{}_{b_i}    (\pi_{ 0 -} )^{A_i}{}_{B_i}     T^{b_i  B_i}    (k)           \, .
\end{eqnarray}
Note that now each component   $T_{+ 1 (-)}^A (k)$  and  $T_{- 1 (-)}^A (k)$ has  $n$  degrees of freedom.

We will consider tensors symmetric in all indices. The other $n - 1$ possibilities are antisymmetric expression in $a_i$  and one of $n -1$  indices from $A_i$.
It means that  each component    $(T)_{+ 1 (-)}^A  (k)$ and    $(T)_{- 1 (-)}^A  (k)$ has  one  degree of freedom.  From now on, we will assume that all tensors are totaly symmetric in all indices,
but  we will not denote symmetrization explicitly. For example  $T_{\pm 1 (-)}^{a_i  A_i}$ means totaly symmetric in all indices $A$.

Using expressions for projectors  we get
\begin{eqnarray}\label{eq:}
T_{\pm 1 (-)}^{a_i  A_i}  (k) =  - \frac{1}{2^n}   \Big(  (\breve{p}_\mp)_{b_j}  q_{B_j} T^{b_i  B_i} (k)  \Big)   \sum_{i = 1}^n     k^{A_i}    \breve{p}_\pm^{a_i}
\equiv      \mathbf{T}_{\mp n}    \sum_{i = 1}^n   k^{A_i}  \breve{p}_\pm^{a_i}      \, .
\end{eqnarray}
The fields  $T_{\pm 1 (-)}^A   (k) $  with helicities $\lambda = \pm 1$  contain   one basic vector $\breve{p}_\pm^a$ and
consequently  relation (\ref{eq:Hln})  is satisfy.

Therefore,  we can write  totally symmetric tensor  in the form
\begin{eqnarray}\label{eq:STpm1r2}
(T_{\pm 1 (-)})^A   (k) =    \sum_{i = 1}^n   k^{A_i}   A_\pm^{a_i}   (k)     \, ,   \qquad
 A_\pm^a =   \mathbf{T}_{\mp n}   \breve{p}_\pm^a  \, ,
\end{eqnarray}

Here  $A_\pm^a $  is  new field. Note that  it is  proportional to $\breve{p}_\pm^a$, as well as fields $V_\pm^a$ in
Eq.(\ref{eq:Vfbv}).  According to scalar product of basic vectors $k_a \breve{p}_\pm^a  = 0$ we can conclude that    $k_a A_\pm^a = 0 $, which means that these solutions are
in Lorentz gauge. We will see that  these  fields  describe electromagnetic interaction.

\subsection{Gauge transformation for arbitrary  rank tensor  with  helicities  $\lambda = \pm 1$ and Maxwell equations }

Using gauge transformations for basic vectors $\delta k^a = 0$ and  $\delta  \breve{p}_\pm^a =   k^a \varepsilon_\pm $ we can obtain  gauge transformations for totally  symmetric
 arbitrary  rank tensor with  helicities $\lambda  = \pm 1$
\begin{eqnarray}\label{eq:Gtrstn}
\delta  (T_{\pm 1 (-)})^A  (k) =  k^A   \omega_\pm  \, , \qquad  \omega_\pm (k) = n  \mathbf{T}_{\mp n}    \varepsilon_\pm      \, .  \qquad
\end{eqnarray}
and for field $A_\pm^a$
\begin{eqnarray}\label{eq:GtrstA}
\delta  A_\pm^a   (k) =  k^a   \tilde{\omega}_\pm  \, , \qquad  \tilde{\omega}_\pm (k) =   \mathbf{T}_{\mp n}    \varepsilon_\pm      \, .  \qquad
\end{eqnarray}

Since $k^a$  and $q^a$ are invariant under space inversion and ${\cal P} \breve{p}_\pm^a = -  \breve{p}_\mp^a$  and  ${\cal P} \varepsilon_\pm = -  \varepsilon_\mp$    we have
${\cal P} (T_{\pm 1 (-)})^A  (k) =  (T_{\mp 1 (-)})^A  (k)$,  ${\cal P}  A^a_\pm (k) =  A^a_\mp (k) $   and
${\cal P}  \omega_\pm  (k) = \omega_\mp (k) $.  So,   according to  (\ref{eq:GSbh})   instead of two components $ (T_{ + 1 (-)})^A    (k )$  and $(T_{-1 (-)})^A    (k )$
we will introduce one  symmetric tensor
\begin{eqnarray}\label{eq:GSv12n}
 T^A   (k ) =   \alpha  (T_{ + 1 (-)})^A    (k )   +  \beta  (T_{-1 (-)})^A    (k )  \, ,  \qquad
\end{eqnarray}
and  instead of two components   $A^a_\pm$   we will introduce one vector  field
\begin{eqnarray}\label{eq:GSv11n}
 A^a  (k ) =   \alpha  A^a_+   (k )   +  \beta  A^a_-   (k )  \, ,  \qquad
\end{eqnarray}
which  are incomplete  irreducible representation of Poincare group with  space inversion.
The representation of the field  $T^A  (k) $ as well as of  $ A^a  (k )$ is two dimensional and consequently both fields  have  two degrees of freedom.
Since the parameters $ \omega_\pm (k) $  and  $ {\tilde \omega}_\pm (k) $  are also connected by parity transformation we can introduce one parameter $\omega (k) = \alpha \omega_+  (k) +    \beta \omega_- (k)$  and  one parameter $ {\tilde \omega} (k) = \alpha  {\tilde \omega}_+  (k) +    \beta  {\tilde \omega}_- (k)$ .

If we go to arbitrary frame  and coordinate  dependent   fields,   $k^a  \to p^a \to    i \partial^a$,  we obtain  expression  for totally  symmetric tensor
$ T^A  (x)$ in terms of vector field  $A^a (x)$ and theirs gauge transformations.    For real field,   according to  (\ref{eq:GSv12n})   and  (\ref{eq:GSv11n})   
\begin{eqnarray}\label{eq:Gthabn}
T^A   (x)  =   \sum_{i = 1}^n    \partial^{A_i}   A^{a_i} (x)      \, .  \qquad
\end{eqnarray}

To obtain gauge transformation we use    (\ref{eq:Gtrst}) and    (\ref{eq:GtrstA})   and obtain
\begin{eqnarray}\label{eq:Gthabdn}
\delta  T^A   (x)   =  \partial^A   \omega (x)      \, .  \qquad
\delta  A^a  (x) =   \partial^a   \omega_1  (x)   \, ,  \qquad   \omega_1 (x) = \frac{1}{n} \omega   (x) \, .
\end{eqnarray}

The field $A^a  (x)$  behaves like  electromagnetic field. It is vector field, it has the same  helicity, the same  gauge transformation  and the same number of degrees of freedom. 
So, this  describes  electromagnetic interaction.

\subsubsection{Antisymmetric tensor with   helicities  $\lambda = \pm 1$   and   Maxwell equations}

We already mentioned  that there are  $n - 1$ possibe   antisymmetric expressions in $a_i$  and one of $n -1$  indices from $A_i$.
It means that  each component    $(T^{An})_{+ 1 (-)}^A  (k)$ and    $(T^{An})_{- 1 (-)}^A  (k)$  has  one  degree of freedom.    
Using expressions for projectors  we get
\begin{eqnarray}\label{eq:TAnaij}
(T^{An})_{\pm 1 (-)}^{a_i a_j A_{i j}}  (k) =  - \frac{1}{2^n}   \Big(  (\breve{p}_\mp)_{b_k}     k_{b_q}  q_{B_{k q}}   T^{b_k b_q B_{k q}} (k)  \Big) 
  \sum_{i, j = 1}^n      k^{A_{i j}}    \Big(  k^{a_i} \breve{p}_\pm^{a_j}     -     k^{a_j}  \breve{p}_\pm^{a_i}      \Big)        \nonumber \\
\equiv      \mathbf{T}_{\mp n}    \sum_{i, j = 1}^n     k^{A_{i j}}    \Big(  k^{a_i} \breve{p}_\pm^{a_j}     -     k^{a_j}  \breve{p}_\pm^{a_i}      \Big)        \, .   \qquad 
\end{eqnarray}
Since $k_a k^a = 0$  and  $k_a  \breve{p}_\pm^a = 0$  there are  constraints  $ k_{a_i}  (T^{An})_{\pm 1 (-)}^{a_i a_j A_{i j}}     (k) = 0 $  and  
 $ k_{A_{i j}} T_(T^{An})_{\pm 1 (-)}^{a_i a_j A_{i j}}     (k) = 0 $.
The fields  $(T^{An})_{\pm 1 (-)}^{a_i a_j A_{i j}}    (k) $  with helicities $\lambda = \pm 1$  contain   one basic vector $\breve{p}_\pm^a$ and
consequently  relation (\ref{eq:Hln})  is satisfy.

Both $k^a $  and  $\Big(  k^{a_i} \breve{p}_\pm^{a_j}     -     k^{a_j}  \breve{p}_\pm^{a_i}      \Big)$ are  gauge invariant.  Consequently  fields  $(T^{An})_{\pm 1 (-)}^{a_i a_j A_{i j}}    (k) $  are   Lorentz invariant and we can use them  as physical fields.  
Instead of two components $ (T^{An}_{ + 1 (-)})^A   (k )$  and $(T^{An}_{-1 (-)})^A   (k )$,  according to   (\ref{eq:GSbh})    we will introduce one  antisymmetric tensor
\begin{eqnarray}\label{eq:GSv12An}
F^A   (k ) =   \alpha  (T^{An}_{ + 1 (-)})^A   (k )   +  \beta  (T^{An}_{-1 (-)})^A    (k )  \, ,  \qquad
\end{eqnarray}
which  is irreducible representation of Poincare group with  space inversion. 
Similarly,  from  (\ref{eq:TAnaij})  we have
\begin{eqnarray}\label{eq:Fabkg}
F^A  (k )  =      k^{A_{i j}}   F^{a_i a_j}  \, ,   \qquad             F^{a_i a_j}  =   k^{a_i}       A^{a_j}  (k) -   k^{a_j}  A^{a_i}   (k)      \,  ,  \qquad    \nonumber \\
A^a  (k) =  \alpha  A_+^a  (k)  + \beta A_-^a   (k)      \, .  \qquad  
\end{eqnarray}

Using    scalar product of basic vectors   we can conclude that   there  are two constraints on field $F^{a b} (k)$
\begin{eqnarray}\label{eq:}
k_a   F^{a b}  (k)  = 0     \,  ,  \qquad    \varepsilon_{a b c d}   k^b    F^{c d} (k) =  0     \, .   \qquad
\end{eqnarray}

If we go to arbitrary frame  and coordinate  dependent   fields,   $k^a  \to p^a \to    i \partial^a$,   in the same way as before  we get
\begin{eqnarray}\label{eq:}
\partial_a   F^{a b}  (x)  = 0     \,  ,  \qquad    \varepsilon_{a b c d}   \partial^b    F^{c d} (x) =  0     \, ,  \qquad
\end{eqnarray}
which are just vacuum  Maxwell equations.
As well as in the previous case the representation of the field  $F^{a b} (x) $  is two dimensional and consequently it  has two degrees of freedom.
We could have expected that,  since it describes   Maxwell equations.

To express  coupling of  electromagnetic fields with   matter (to a conserved current) we must solve second equation which produces
\begin{eqnarray}\label{eq:FabAn}
F_{a b} (x)   = \partial_a A_b (x) -   \partial_b A_a  (x)   \, ,  \qquad
\end{eqnarray}
which   also  follows from   (\ref{eq:Fabkg}).

Note that   $A_a (x)$ is gauge field since $ F_{a b} (x)$  is invariant under transformation $\delta A_a (x) =  \partial_a \Omega$.  Then interaction with a matter can be express as in  (\ref{eq:Inm}).

\subsection{Symmetric tensor with   helicities  $\lambda = \pm 2$   and  free Einstein   equations}

  Similarlly as in the previous  subsections   we can consider fields with two    projectors $( \pi_\pm  )^a{}_b$    and $n - 2$  projectors   $( \pi_{ 0 -} )^a{}_b$   
(gauge  invariant     part of $( \pi_{ 0 } )^a{}_b$).    
  
Using expressions for projectors  we  obtain 
\begin{eqnarray}\label{eq:}
(T^{Sy})_{\pm 2 (-)}^{a_i a_j A_{i j}}  (k)      
=   \frac{1}{2^n}   \Big(  (\breve{p}_\mp)_{b_k}     (\breve{p}_\mp)_{b_q}  q_{B_{k q}}   T^{b_k b_q B_{k q}} (k)  \Big) 
  \sum_{i, j  = 1}^n      k^{A_{i j }}      \breve{p}_\pm^{a_i}   \breve{p}_\pm^{a_j}       
\equiv      \mathbf{T}_{\mp n}   \sum_{i, j  = 1}^n    k^{A_{i j }}        \breve{p}_\pm^{a_i}   \breve{p}_\pm^{a_j}          \, .  \qquad      
\end{eqnarray}

It is proportional to the tensor (\ref{eq:Hhpm2}) 
\begin{eqnarray}\label{eq:Hhpm2p}
T_{\pm n}^{a b} (k) =       \frac{1}{4}   (\breve{p}_\mp)_c   (\breve{p}_\mp)_d    T^{c d} (k)  \breve{p}_\pm^a      \breve{p}_\pm^b       \, ,  \qquad
\end{eqnarray}
so that 
\begin{eqnarray}\label{eq:Hhpm2pn}
(T^{Sy})_{\pm 2 (-)}^{a_i a_j A_{i j}}  (k)     \sim   \sum_{i, j  = 1}^n     k^{A_{i j }}     T_{\pm 2}^{a_i a_j } (k)         \,   .   \qquad
\end{eqnarray}

If we go to arbitrary frame  and coordinate  dependent   fields,   $k^a  \to p^a \to    i \partial^a$,   in the same way as before  we get 
\begin{eqnarray}\label{eq:Hhpm3p}
H^{a_i a_j A_{i j}}  (x)     \sim   \sum_{i, j  = 1}^n    \partial^{A_{i j }}   h^{a_i a_j } (x)      \,   .   \qquad
\end{eqnarray}

Therefore,  corresponding field is    $n - 2$ derivative  of the field $h^{a b}$  which,  as we showed earlier,   describe  Einstein   equations in weak field approximation.

\section{ Conclusion   }
\setcounter{equation}{0}

Our fundamental assumption is that all field equations for an arbitrary symmetry group can be derived from the corresponding Casimir eigenvalue equations. In this work, we explicitly confirm this statement for the case of the massless Poincaré group.

Through our systematic approach, we have verified Weinberg's observation \cite{W} that Lorentz invariance is violated by terms taking the form of local gauge transformations. 
The principle  field equations provide a simpler confirmation of Weinberg's approach by working directly with fields rather than creation and annihilation operators. Indeed, these equations allow for a more straightforward derivation of the corresponding actions, which play a crucial role in quantization.  
The origin of Weinberg's observation   lies in the non-commutativity of the Pauli-Lubanski vector components. We have rigorously verified these results for tensor fields of arbitrary rank.

The introduction of the projectors   is crucial because their diagonalization of the characteristic polynomial allows us to solve  the corresponding determinants of arbitrary order and determine the spectrum of all helicities.

We defined the basis vectors such that the helicity  $\lambda$   takes a simple form in terms of these vectors.   Specifically, we have  $\lambda = n_+ -  n_-$,  where $n_\pm$  denotes the number of basis vectors $\breve{p}_\pm^a$.  
We emphasize that helicity plays a fundamental role in determining both the field equations for irreducible components of a given field   $\Psi^A$   and the number of their physical degrees of freedom.  
Importantly, the same physical theory can be described using different field representations.
For instance, as demonstrated in Eq.(\ref{eq:Hln})  all representations  with helicities $\lambda = \pm 1$  (including vector fields, symmetric and antisymmetric second-rank tensors  and arbitrary rank tensors )
exhibit one additional vector   $\breve{p}^a_\pm$  compared to     $\breve{p}^a_\mp$ 
and consistently describe electromagnetic interactions.

Similarly, representations with helicities   $\lambda = \pm 2$ (such as symmetric second-rank tensors and fourth-rank tensors with appropriate symmetry properties)  
possess two additional vectors    $\breve{p}^a_\pm$  over  $\breve{p}^a_\pm$    and   naturally describe  gravitation  interaction.

These results challenge the conventional view that only the highest helicity of a given tensor field is physically significant. Instead, we demonstrate that all helicity states are physically relevant. 
While different field representations with the same helicity can describe identical physics, formulating the field equations in terms of certain representations may prove non-trivial in practice.

While the primary contribution of this work is the demonstration that the 
principle field equations   (\ref{eq:EMmsc2})  generate all known free field theory equations, our results also represent a significant advancement beyond previous studies of highest helicity fields 
 \cite{FP, SH, Fr,dWF}.   
In all existing literature, only free relativistic fields with the highest helicity,  based  to Fierz-Pauli constraints, have been studied. This represents a  particular case within present  approach, with  totally symmetric tensor fields. 
Unlike previous studies limited to highest-helicity fields, this approach also covers lower-helicity 
not totally symmetric  fields.  These fields admit a clear physical interpretation, as detailed in Sections 7, 8, and 12.

A crucial logical progression emerges in our framework:
{\it Covariant   helicity definition   $\to$  little group invariance  $\to$  local gauge transformations $\to$   equations of motion  $\to$  actions  for massless fields}.  

This sequence reveals that the requirement of Lorentz invariance necessarily leads to the existence of local gauge transformations for massless fields.

\end{document}